\documentclass[journal]{IEEEtran}
\IEEEoverridecommandlockouts
\usepackage{cite}
\usepackage{bm}
\usepackage{mathrsfs}
\usepackage{algorithmic}
\usepackage{multirow}
\usepackage{makecell}
\usepackage{array}
\ifCLASSOPTIONcompsoc
\usepackage[caption=false,font=normalsize,labelfont=sf,textfont=sf]{\part{title}subfig}
\else
\usepackage{subcaption}
\fi
\usepackage{url}
\usepackage{graphicx,amsmath,amssymb,amsfonts}
\usepackage{algorithmic,algorithm}
\usepackage{stfloats}
\allowdisplaybreaks[3]

\newtheorem{theorem}{\textbf{Theorem}}

\newtheorem{lemma}{\textbf{Lemma}}

\newtheorem{corollary}{\textbf{Corollary}}

\newtheorem{definition}{\textbf{Definition}}
\newtheorem{assumption}{\textbf{Assumption}}

\makeatletter

\newcommand{\Rmnum}[1]{\expandafter\@slowromancap\romannumeral #1@}

\makeatother
\begin{document}

\title{{\fontsize{23 pt}{\baselineskip}
			\selectfont Task-Agnostic Semantic Communications Relying on Information Bottleneck and Federated Meta-Learning} 
	}

\author{\IEEEauthorblockN{
		Hao~Wei, 
		Wen~Wang, 
		Wanli~Ni, 
		Wenjun~Xu,~\IEEEmembership{Senior Member,~IEEE}, 
	    Yongming~Huang,~\IEEEmembership{Fellow,~IEEE},
		Dusit~Niyato,~\IEEEmembership{Fellow,~IEEE},
	    and Ping~Zhang, ~\IEEEmembership{\!\!Fellow,~IEEE}
		}

	\thanks{H. Wei, W. Xu, and P. Zhang are with the State Key Laboratory of Networking and Switching Technology, Beijing University of Posts and Telecommunications, Beijing 100876, China. W. Xu and P. Zhang are also with the Department of Mathematics and Theories, Peng Cheng Laboratory, Shenzhen 518066, China (e-mail: hao.wei@bupt.edu.cn; wjxu@bupt.edu.cn; pzhang@bupt.edu.cn). (\emph{Corresponding author: Wenjun Xu}). }
	
	\thanks{W. Wang and Y. Huang are with the School of Information Science and Engineering, and the National Mobile Communications Research Laboratory, Southeast University, Nanjing 210096,	China, and also with the Pervasive Communications Center, Purple Mountain	Laboratories, Nanjing 211111, China (e-mail: wangwen@pmlabs.com.cn; huangym@seu.edu.cn).}
	
	\thanks{W. Ni is with the Department of Electronic Engineering, Tsinghua University, Beijing 100084, China (e-mail: niwanli@tsinghua.edu.cn).}

	\thanks{D. Niyato is with School of Computer Science and Engineering, Nanyang Technological University, Singapore 117583 (e-mail: dniyato@ntu.edu.sg).} 

}

\maketitle
\begin{abstract}
As a paradigm shift towards pervasive intelligence, semantic communication 
(SemCom) has shown great potentials to improve communication efficiency and provide user-centric services by delivering task-oriented semantic meanings.
However, the exponential growth in connected devices, data volumes, and communication demands presents significant challenges for practical SemCom design, particularly in resource-constrained wireless networks.
In this work, we first propose a task-agnostic SemCom (TASC) framework that can handle diverse tasks with multiple modalities.
Aiming to explore the interplay between communications and intelligent tasks from the information-theoretical perspective, we leverage information bottleneck (IB) theory and propose a distributed multimodal IB (DMIB) principle to learn minimal and sufficient unimodal and multimodal information effectively by discarding redundancy
while preserving task-related information.
To further reduce the communication overhead, we develop an adaptive semantic feature transmission method under dynamic channel conditions.
Then, TASC is trained based on federated meta-learning (FML) for rapid adaptation and generalization in wireless networks.
To gain deep insights, we rigorously conduct theoretical analysis and devise resource management to accelerate convergence while minimizing the training latency and energy consumption.
Moreover, we develop a joint user selection and resource allocation algorithm to address the non-convex problem with theoretical guarantees.
Extensive simulation results validate the effectiveness and superiority of the proposed TASC compared to baselines.

\end{abstract}

\section{Introduction}
With the prosperous advancements of artificial intelligence (AI), semantic communication (SemCom) has emerged as a new promising candidate for next-generation communications\cite{Beyond}. This innovative paradigm focuses on extracting and transmitting the underlying meaning of source data behind digital bits, enhancing communication efficiency and intelligence.
As we enter the era of ``Internet of Everything'', wireless networks are evolving vigorously to accommodate various kinds of intelligent interactive applications, such as virtual reality, automatic driving, and digital twining\cite{Engineering,You2025When}.
However, the rapid development of emerging technologies has led to escalating demands for increasingly complex, diverse, and intelligent transmission. 
Concurrently, the proliferation of connected devices has resulted in a substantial surge 
in wireless traffic and imposes significant pressure on network capacity, presenting critical challenges for SemCom\cite{Survey}.

On the one hand, deep learning (DL)-based SemCom 
predominantly relies on manually designs that adopt state-of-the-art deep neural networks (DNNs). 
Less efforts have been focused on the interplay between wireless communication and intelligent tasks, hindering further improvement in SemCom\cite{TWC-Explain}.
On the other hand, when the communication environment changes, resulting in variations of tasks and data distributions, the semantic transceiver typically requires redesigning, updating, and performing switching, which in turn leads to substantial storage and computational inefficiency.
This process limits its ability to seamlessly adapt to diverse tasks and brand-new data, highlighting its restricted generalization and self-adaptation capabilities\cite{FML-Survey}.
Moreover, 
transmitting user-collected data to a centralized server for semantic learning is commonly infeasible due to constrained communication resources and privacy concerns\cite{TMC-FedSL}.
Although the local semantic training presents a potential alternative, it is impractical
as SemCom inherently depends on the coordinated deployment of transceiver networks at both the transmitter and receiver\cite{FL-train}.
Therefore, it is imperative to develop a universal
SemCom framework that can effectively address the above challenges.

To support large-scale connectivity, multiuser SemCom has been investigated recently. For example, Wang \emph{et al.}\cite{DistributedJSCC} proposed a distributed DL-based joint source-channel coding (JSCC) scheme for wireless image transmission.
Zhang \emph{et al.}\cite{VTM} developed a cooperative multiuser SemCom to curtail the data traffic in Internet of Vehicles.
Then, Weng \emph{et al.}\cite{MIMO-speech} presented a semantic multiple-input multiple-output (MIMO) system for speech-to-text transmission.
Although the aforementioned SemCom systems have demonstrated satisfactory performance in specific scenarios, they are limited to handling a single unimodal task.
Afterwards, Wang \emph{et al.}\cite{SPL-multi-task} proposed a multi-task learning network for JSCC to handle the image detection and segmentation tasks.
Xie \emph{et al.}\cite{WCL-Multi-User,JSAC-Multi-User}
proposed multiuser SemCom systems for transmitting and integrating text and image data to accomplish visual question answering (VQA) task.
Zhang \emph{et al.}\cite{Unified-SemCom} investigated a unified multi-task SemCom
system for multimodal data.
However, 
these studies\cite{DistributedJSCC,VTM,MIMO-speech,SPL-multi-task,WCL-Multi-User,JSAC-Multi-User,Unified-SemCom} rely on stacking advanced DNNs to enhance performance, lacking a systematic and comprehensive theoretical framework to exploit the interaction between communication and intelligent task.
Moreover, models have to be updated once the task/data distributions change, which leads to massive gradient transmission to adapt to new environments.
Besides, they adopt centralized learning by default and ignore the practical training and deployment in wireless networks.

Emerged as an effective approach to deployment, federated learning (FL) has been exploited to collaboratively train a model using locally available data while leveraging a centralized server for global aggregation\cite{FL-Survey}.
Specifically, Chen \emph{et al.}\cite{Chen-TWC} proposed a joint learning and communication framework based on FL.
By introducing FL into SemCom,
Tong \emph{et al.}\cite{FL-speech} considered an FL-based SemCom system for audio transmission and recovery. 
Then, Wei \emph{et al.}\cite{CL-Semcom} presented a novel federated semantic learning framework for knowledge graph generation.
Wang \emph{et al.}\cite{FL-constrasive} designed a federated contrastive
learning system to support personalized SemCom.
Xu \emph{et al.}\cite{FL-distillation} leveraged group feature distillation to train edge models and enhanced global image SemCom performance.
Nevertheless, 
FL usually suffers from statistical challenges (i.e., 
personalized and heterogeneous characteristics) and systematic challenges (i.e., limitation of storage, computation, and communication capacities), resulting in poor adaptation and generalization\cite{FML-fast}.

As a synergy of FL and meta-learning (also known as \emph{learning to learn}), federated meta-learning (FML) has stimulated significant interest\cite{FML-Survey}. In FML, users collaboratively learn an initial shared model under the coordination of a central server, enabling both existing and new users to efficiently adapt the learned model to their local datasets with only one or a few gradient descent steps. 
Notably, FML retains the advantages of FL, while providing a more personalized model for each device to quickly accommodate task and data heterogeneity (namely task/model agnostic)\cite{FML-agnostic}. Consequently, FML is a promising solution to address the statistical and systematic challenges with fast adaptation and good generalization\cite{FML-Personalized}.

Despite the auspicious benefits of FML, we still encounter the following challenges.
Firstly, distributed coding is an important problem in information theory and communication. An optimal semantic representation should retain \emph{sufficient} information for accurate inference while keeping \emph{minimal} information without redundancy and semantic noise\cite{IB-JSAC}. There lacks a systematic way to deal with such a tradeoff between the informativeness of distributed multimodal features and their impact on tasks, especially under dynamic channel conditions.
Secondly, rather than retraining task-specific models separately, 
it is crucial to develop a task-agnostic semantic transceiver network capable of supporting diverse tasks.
A key problem lies in effectively exploring unimodal and multimodal representations while ensuring consistency and complementarity across modalities.
Thirdly, 
a reasonable multiuser SemCom requires the transceiver networks to be deployed at the semantic users (SUs) and the base station (BS), respectively\cite{CL-Semcom}. 
In this case, integrating FML into SemCom to enable rapid adaptation to new environments with brand-new users and data remains an open problem. 
Meanwhile, rigorous performance analysis and efficient resource management are essential for ensuring the sustainability and stability of the system\cite{Energy-Efficient}.

To overcome the above challenges, we propose a task-agnostic semantic communication (TASC) framework that can handle various intelligent tasks.
The primary contributions of this paper are summarized as follows:
\begin{itemize}
	
\item We propose a novel TASC framework,
where multiple task-oriented semantic transmitter networks at SUs extract and transmit unimodal semantic information, which is then processed by a shared task-agnostic semantic receiver network at the BS to generate multimodal representation for multiple diverse tasks.

\item We introduce the information bottleneck (IB) principle into TASC to facilitate unimodal and multimodal semantic coding.
Two variants of distributed multimodal IB (DMIB) are proposed: 1) univocal DMIB, which learns minimal sufficient unimodal representations without redundancy while reserving task-relevant information, and
2) syncretic DMIB, which ensures task-related multimodal representations by filtering out semantic noise.
Furthermore, to reduce the communication overhead, we develop an adaptive semantic feature transmission scheme by dynamically adjusting the transmission dimensions under varying channel conditions.

\item Then, we use the FML to train TASC 
to achieve rapid adaptation and generalization.
To gain deep insights, we conduct convergence analysis of TASC under non-convex settings.
We then perform resource allocation to capture the tradeoff among convergence speed, training latency, and energy cost. 
Finally, a joint optimization algorithm is proposed to effectively solve the non-convex problem.

\item Through extensive experiments, TASC obtains satisfactory performance under dynamic channel conditions compared to the task-specific SemCom.
By introducing the IB principle, TASC achieves a better rate-distortion tradeoff.
We also validate the superiority of TASC over baselines in terms of convergence speed, training latency, and energy efficiency.
Furthermore, 
leveraging the benefits of meta-learning, 
TASC demonstrates rapid adaptation capabilities by achieving about a 59\% decrease in terms of test loss than FL through few gradient descent iterations.

\end{itemize}

\begin{figure*}[t]
	\includegraphics[width=0.90\textwidth]{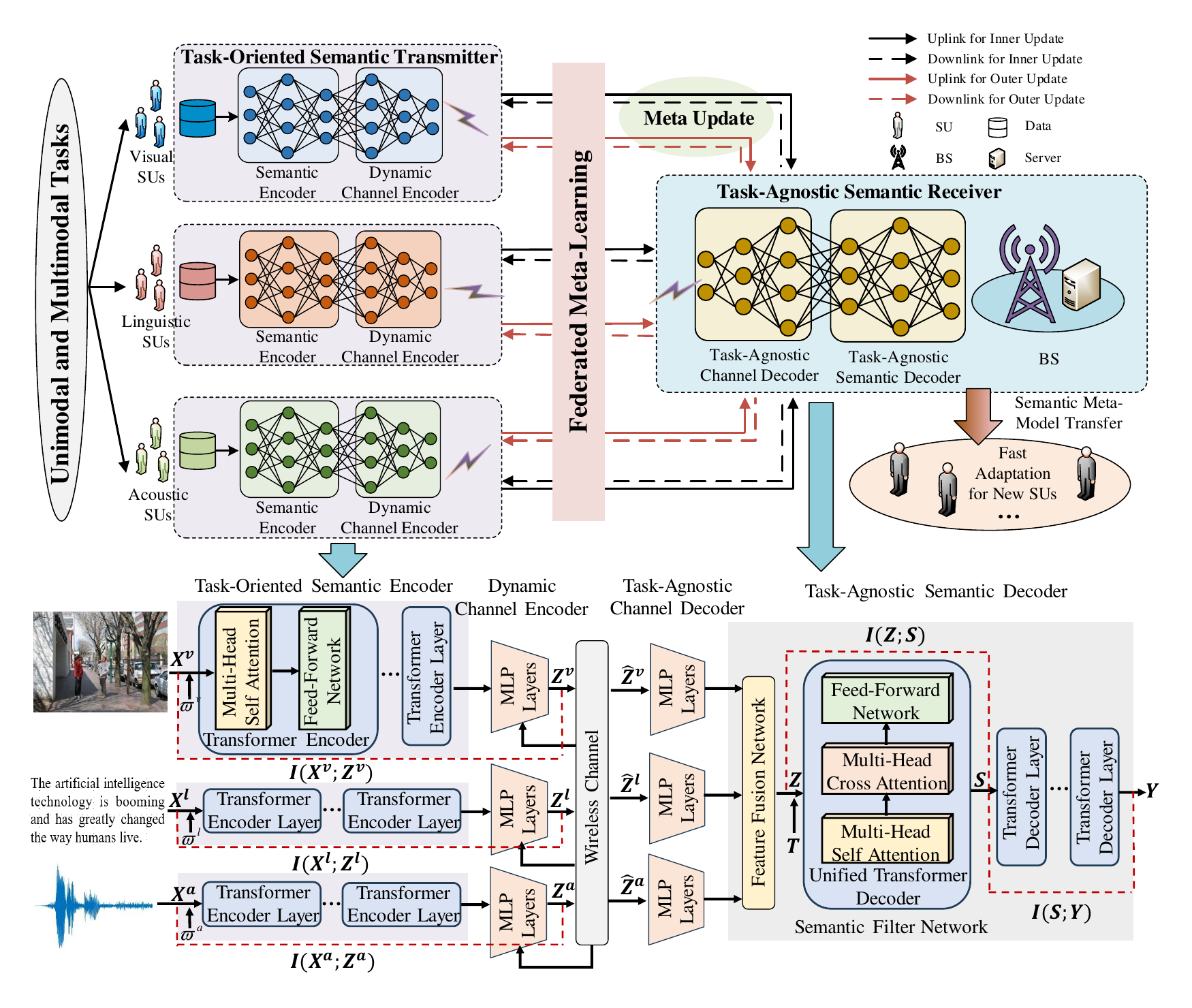} 
	\centering 
	\caption{The diagram of the proposed TASC framework over wireless networks.}  
	\label{TASC-framework}  
\end{figure*}

\section{TASC System Design}

In this section, we first introduce the TASC system driven by the DMIB principle 
Then, we propose an adaptive semantic feature transmission scheme under dynamic channel conditions.
Finally, the task-agnostic semantic-channel coding structure is depicted.


\subsection{Semantic Communication Model}

As illustrated in Fig. \ref{TASC-framework}, we consider a wireless multimodal multiuser SemCom network, where three modalities are considered, i.e., visual ($v$), linguistic ($l$), and acoustic ($a$) modalities.
Each modality $m\in\{v,l,a\}$ comprises
a set $\mathcal K^m=\{1,\ldots,K^m\}$ of $K^m$ SUs. 
These SUs employ SemCom techniques to transmit the semantic meaning behind the source datas to a BS in a task-oriented manner. 
Each SU 
is connected to the BS equipped with an edge server.
The local dataset of the $k$-th SU with modality $m$ is defined as $\mathcal{D}_k^m$ with size ${D_k^m}$, 
and thus the whole dataset for modality $m$ is denoted as $\mathcal{D}^m=\bigcup_{k \in \mathcal{K}^m} \mathcal{D}_k^m$. 
Let $(\boldsymbol{x}_{k}^{m}, y_k)$ 
denote the sample points which follow an underlying distribution $P_k^m$ for modality $m$,
where $\boldsymbol{x}_{k}^{m}\in\mathbb{R}^{q_k^m}$ represents the input data samples with $q_k^m$ being the dimension of modality $m$ for SU $k$, and $y_{k}$ denotes the true label.
As shown in Fig. \ref{TASC-framework}, each SU is deployed with a semantic transmitter including a semantic encoder and a channel encoder, both of which are represented by DNNs. 
Then, the encoding process is formulated as
\setlength{\abovedisplayskip}{3pt}
\setlength{\belowdisplayskip}{3pt}
\begin{eqnarray} 
	~~~\boldsymbol z_{k}^{m}={\mathcal {ST}_m(\boldsymbol x_{k}^{m};{{{\boldsymbol{\theta}_{k}^m}}})}, ~\forall k \in \mathcal{K}^m, m\in\{v,l,a\},
\end{eqnarray}
where ${\mathcal {ST}_m(\cdot;{{{\boldsymbol{\theta}_{k}^m}}})}$ denotes the semantic transmitter parameterized by $\boldsymbol{\theta}_{k}^m$ of the $k$-th SU for processing modality $m$. 
It is worth noting that the orthogonal frequency division multiplexing (OFDM) signal processing is utilized in this work to convert semantic data into transmitted signals. As this operation can be formulated as a series of linear transformations (i.e., matrix vector multiplications), it is seamlessly integrated into the channel coding process within the DNNs\cite{OFDM-JSCC}. 
Afterwards, the transmitted signal $\boldsymbol z_{k}^{m}\in\mathbb{C}^{c_k^m}$ with signal length $c_k^m$ passes a wireless channel. 
The received noise-corrupted signal from the $k$-th SU at the BS is given by
\begin{eqnarray} 
	~~~~\boldsymbol {\hat z}_{k}^{m}=h_k^m\boldsymbol z_{k}^{m}+ \boldsymbol n_{k}^{m}, ~\forall k \in \mathcal{K}^m, m\in\{v,l,a\},
\end{eqnarray}
where $h_k^m\in\mathbb{C}$ denotes the channel gain, and $ \boldsymbol n_{k}^{m}\sim\mathcal{CN}(\mathbf{0},\;\sigma^2\mathbf{I})$ represents the independent and identically distributed Gaussian noise with zero mean and variance $\sigma^2$.

Subsequently, the received signal can be recovered by the  channel decoder and then fed into the semantic decoder to directly produce the inference results ${\boldsymbol{\hat y}}$, expressed as
\begin{eqnarray} 
	~~~~{\boldsymbol{\hat y}}={\mathcal {SR}(\boldsymbol {\hat z}_{1}^{v},\boldsymbol {\hat z}_{2}^{v},\ldots,\boldsymbol {\hat z}_{k}^{l},\ldots,\boldsymbol {\hat z}_{K^a}^{a};{{\boldsymbol{\theta}_B}})},
\end{eqnarray}
where $\mathcal {SR}(\cdot;{{\boldsymbol{\theta}_B}})$ indicates the shared BS-side semantic receiver parameterized by $\boldsymbol{\theta}_B$.

\subsection{Distributed Multimodal Information Bottleneck}

Given the defined SemCom model, the variables are characterized by the following Markov chain\cite{edgeDIB}:
\begin{align}
\label{Markov}
\!\!	Y_k \leftrightarrow	X_k^m \leftrightarrow  Z_k^m \leftrightarrow  Z_k \leftrightarrow  S_k,  \!~\forall k \in \mathcal{K}^m, m\in\{v,l,a\},
\end{align}
which satisfies $p({\boldsymbol{s}_k}, {\boldsymbol{z}_k},{\boldsymbol{ z}_k^m}, {\boldsymbol{ x}_k^m} |\boldsymbol{y}_k)=  p(\boldsymbol{s}_k|\boldsymbol{z}_k) p({\boldsymbol{ z}_k}| {\boldsymbol{ z}_k^m})$ $p({\boldsymbol{ z}_k^m}|{\boldsymbol{x}_k^m}) p({\boldsymbol{x}_k^m}|{\boldsymbol{ y}_k}) $. Note that $Y_k, X_k^m, Z_k^m, Z_k, S_k$ are random variables, and $ {\boldsymbol y}_k, {\boldsymbol{ x}_k^m}, {\boldsymbol{ z}_k^m}, {\boldsymbol{z}_k} , {\boldsymbol{s}_k}$ are multiple instances of corresponding random variables. 
Firstly, we revisit the traditional IB and assume the Markov chain $	Y_k \leftrightarrow	X_k \leftrightarrow  Z_k $.
To achieve efficient task-oriented SemCom,
the semantic feature $Z_k$ should be extracted and compressed in a concise representation by discarding the task-irrelevant information from the observation $X_k$, and preserved only informative task-related information about the target variable $Y_k$.
The IB principle aims to minimize the mutual information between
$Z_k$ and $X_k$ while maximizing the mutual information between $Z_k$ and $Y_k$\cite{IB-JSAC}.
The objective of IB is formulated by
\begin{align}
	\label{IB}
\!\!\!	\mathcal{L}_k^{\text{IB}}\!\!=\!\zeta \underbrace{I(X_k; Z_k)}_{\text {Rate}}\underbrace{-I(Y_k; Z_k)}_{\text{Distortion}}  
	\equiv \zeta I(X_k; Z_k)\!+\!H(Y_k|Z_k),\!\!\!\!
\end{align}
where $\zeta$ is a weight factor and $H(Y_k)$ is a constant that can be eliminated as
$I(Y_k; Z_k)=H(Y_k)-H(Y_k|Z_k)$.

\subsubsection{Univocal DMIB}
Unlike the traditional IB, our system processes the input comprising multiple unimodal representations $X^v$, $X^l$, and $X^a$, which may contain disturbed task-unrelated information, hindering the effective exploration of complementary information among modalities.
Moreover, the extracted features should retain \emph{minimal} information by dislodging redundant elements to achieve efficient transmission.
Therefore, we propose a new univocal DMIB (U-DMIB), which focuses on discriminative unimodal semantic features with \emph{minimal} and \emph{sufficient} representations to eliminate task-irrelevant information prior to fusion and align their encoded distributions by independently applying the IB principle to each modality.
The objective function is formulated as
\begin{align}
\label{U-DMIB}
\mathcal{L}_k^{\text{U-DMIB}}=\sum\nolimits_{m}[\zeta I(X_k^m;  Z_k^m)+ H(Y_k| Z_k^m)].
\end{align}
For the rate term in (\ref{U-DMIB}), it is expressed as
\begin{align}
\label{U-DMIB-KL}
I(X_k^m; Z_k^m)=D_\text{KL}{(p_{\boldsymbol{\theta}_k^m}(\boldsymbol{ z}_k^m|\boldsymbol{x}_k^m)||p{(\boldsymbol {z}_k^m)})}.
\end{align}
Given the joint distribution $p(\boldsymbol{x}_k^m,\boldsymbol{y}_k)$ and the Markov chain in (\ref{Markov}), $p{(\boldsymbol { z}_k^m)}$ depends on the $p_{\boldsymbol{\theta}_k^m}(\boldsymbol{ z}_k^m|\boldsymbol{x}_k^m)$, expressed as
\begin{align}
\label{U-DMIB-pz}
p{(\boldsymbol { z}_k^m)}=\smallint  p\left(\boldsymbol{x}_k^m\right) p_{\boldsymbol{\theta}_k^m}\left(\boldsymbol{z}_k^m | \boldsymbol x_k^m\right) d \boldsymbol x_k^m.
\end{align}
However, this distribution is intractable due to the high-dimensional integral.
To overcome this issue, we introduce variational distribution $q{(\boldsymbol { z}_k^m)}$ to approximate distribution $p{(\boldsymbol { z}_k^m)}$\cite{VIB}.
Due to the non-negativity of the Kullback-Leibler (KL) divergence, it is obtained that
$D_\text{KL}(p{(\boldsymbol { z}_k^m)}||q{(\boldsymbol {z}_k^m)})$, and thus the variational upper bound is given by
\begin{align}
	\label{U-DMIB-varia-rate}
I(X_k^m; Z_k^m)\leq D_\text{KL}{(p_{\boldsymbol{\theta}_k^m}(\boldsymbol{ z}_k^m|\boldsymbol{x}_k^m)||q{(\boldsymbol {z}_k^m)})}.
\end{align}
For the distortion term in (\ref{U-DMIB}), $p(\boldsymbol{y}_k|\boldsymbol{ z}_k^m)$ is typically impossible to compute due to the high dimensionality, denoted as
\begin{align}
\label{U-DMIB-pzy}
p(\boldsymbol{y}_k|\boldsymbol{ z}_k^m)=\frac{\int p\left(\boldsymbol{x}_k^m,\boldsymbol{ y}_k\right) p_{\boldsymbol{\theta}_k^m}\left(\boldsymbol{z}_k^m | \boldsymbol x_k^m\right) d \boldsymbol x_k^m}{p{(\boldsymbol { z}_k^m)}}.
\end{align}
Similarly, we introduce $q_{\boldsymbol{ \theta}_B}(\boldsymbol{y}_k|\boldsymbol{ z}_k^m)$ as the variational distribution to approximate the true distribution $p(\boldsymbol{y}_k|\boldsymbol{ z}_k^m)$.
Therefore, the objective function of variational U-DMIB (U-VDMIB) in (\ref{U-DMIB}) is recast as
\begin{align}
	\nonumber
	\mathcal{L}_k^{\text{U-VDMIB}}=&\sum_{m}[\mathbf{E}_{p({\boldsymbol x}_k^m, \boldsymbol{y}_k)}\{ \zeta D_\text{KL}{(p_{\boldsymbol{\theta}_k^m}(\boldsymbol{z}_k^m|\boldsymbol{x}_k^m)||q{(\boldsymbol { z}_k^m)})} \\ \label{U-VDMIB}
	&	+ \mathbf{E}_{\boldsymbol{p}_{\boldsymbol{\boldsymbol { \theta } _ { k }}}\left(\boldsymbol {z}_k^m | \boldsymbol{x}_k^m\right)}[-\log q_{\boldsymbol{ \theta}_B}(\boldsymbol{y}_k|\boldsymbol{ z}_k^m)]
	\}].
\end{align}

Leveraging the capabilities of deep learning, the distributions $p_{\boldsymbol{\theta}_k^m}(\boldsymbol{ z}_k^m|\boldsymbol{x}_k^m)$ and $q_{\boldsymbol{ \theta}_B}(\boldsymbol{y}_k|\boldsymbol{ z}_k^m)$
can be parameterized through DNNs with $\boldsymbol{ \theta}_k^m$ and $\boldsymbol{ \theta}_B$.
Generally, the multivariate Gaussian distribution is commonly used to characterize 
$p_{\boldsymbol{\theta}_k^m}(\boldsymbol{ z}_k^m|\boldsymbol{x}_k^m)$, i.e., $p_{\boldsymbol{\theta}_k^m}(\boldsymbol{ z}_k^m|\boldsymbol{x}_k^m)=\mathcal{N}(\boldsymbol{ z}_k^m|\boldsymbol{\mu}_k^m, \boldsymbol{\Sigma}_k^m)$,
where $\boldsymbol{\mu}_k^m$ and $\boldsymbol{\Sigma}_k^m$ are the mean vector and the covariance matrix learned by the DNNs $\boldsymbol{\mu}_{z_k^m}(\boldsymbol{x}_k^m;\boldsymbol{\theta}_k^m)$ and $\boldsymbol{\Sigma}_{z_k^m}(\boldsymbol{x}_k^m;\boldsymbol{\theta}_k^m)$, respectively.
Besides, the approximated marginal distribution $q{(\boldsymbol { z}_k^m)}$ can be treated as a standard centered isotropic Gaussian distribution
$\mathcal{N}(\boldsymbol { z}_k^m|\mathbf{0},\mathbf{I})$.
To optimize the objective in (\ref{U-VDMIB}) via the stochastic gradient descent (SGD),
the reparameterization trick is adopted to sample $\boldsymbol { z}_k^m$ from $p_{\boldsymbol{\theta}_k^m}(\boldsymbol{ z}_k^m|\boldsymbol{x}_k^m)$, i.e., $\boldsymbol { z}_k^m=\boldsymbol{\mu}_k^m+ \boldsymbol{\Sigma}_k^m\odot\boldsymbol{\iota }_k^m$, where $\boldsymbol{\iota }_k^m\sim\mathcal{N}(\mathbf{0},\mathbf{I})$.
Then, by employing Monte Carlo sampling, we can attain an unbiased estimate of the gradient, enabling the optimization in (\ref{U-VDMIB}). 
Given a mini-batch of data $\{(\boldsymbol{x}_{k}^{m,i},\boldsymbol y_k^i)\}_{i=1}^B$ at SU $k$ for modality $m$, the empirical estimation of the U-VDMIB is expressed as
\begin{align}
	\nonumber
	\mathcal{L}_k^{\text{U-VDMIB}}\simeq&\frac{1}{B}\sum\nolimits_{i=1}^{B}\sum\nolimits_{m}[ \zeta D_\text{KL}\left({\mathcal{N}(\boldsymbol{\mu}_k^{m,i}, \boldsymbol{\Sigma}_k^{m,i})||\mathcal{N}(\mathbf{0},\mathbf{I})}\right) \\ \label{U-VDMIB-emp}
	& -\log q_{\boldsymbol{ \theta}_B}(\boldsymbol{y}_k^i|\boldsymbol{ z}_k^{m,i})].
\end{align}

\subsubsection{Syncretic DMIB}
Different from the above U-DMIB, a key challenge is to ensure that the learned multimodal representation retains sufficient task-related information without semantic noise from different unimodal semantic representations. 
One intuitive approach is to first integrate these unimodal representations into a multimodal fusion representation and subsequently regularize it using the IB principle. 
We refer this approach as syncretic DMIB (S-DMIB), denoted by
\begin{align}
	\label{S-DMIB}
	\mathcal{L}_k^{\text{S-DMIB}}=\zeta I(Z_k;  S_k)+ H(Y_k| S_k),
\end{align}
where $S_k$ is the filtered multimodal representation by the feature filter in the task-agnostic semantic decoder, and $Z_k={\boldsymbol{\hat Z}_k^v}\oplus{\boldsymbol{\hat Z}_k^l}\oplus{\boldsymbol{\hat Z}_k^a}$ is the concatenated features.
Afterwards, we apply IB to reduce redundancy and filter out semantic noisy information from the complex and high-dimensional multimodal representation while preserving sufficient task-relevant information.
Based on the deduction of U-VDMIB, the objective of variational S-DMIB is given by
\begin{align}
	\nonumber
	\mathcal{L}_k^{\text{S-VDMIB}}\simeq&\frac{1}{B}\sum\nolimits_{i=1}^{B}[ \zeta D_\text{KL}\left({\mathcal{N}(\boldsymbol{\mu}_k^{i}, \boldsymbol{\Sigma}_k^{i})||\mathcal{N}(\mathbf{0},\mathbf{I})}\right) \\ \label{S-VDMIB-emp}
	& -\log q_{\boldsymbol{ \theta}_B}(\boldsymbol{y}_k^i|\boldsymbol{ s}_k^{i})],
\end{align}
where $\boldsymbol{\mu}_k=\boldsymbol{\mu}_{s}(\boldsymbol{z}_k;\boldsymbol{\theta}_B^{\mu})$ and $ \boldsymbol{\Sigma}_k=\boldsymbol{\Sigma}_{s}(\boldsymbol{z}_k;\boldsymbol{\theta}_B^{\Sigma})$ are learned by the DNNs $\boldsymbol{\theta}_B^{\mu}$ and $\boldsymbol{\theta}_B^{\Sigma}$, respectively.

\subsubsection{Confluent DMIB}
By combining the advantages of U-DMIB and S-DMIB, we develop the confluent DMIB (C-DMIB), expressed by
\begin{align}
	\nonumber
	\mathcal{L}_k^{\text{C-DMIB}}=&\zeta I(Z_k;  S_k)+ H(Y_k| S_k)\\	\label{C-DMIB}
	&+\sum\nolimits_{m}[\zeta I(X_k^m;  Z_k^m)+ H(Y_k| Z_k^m)].
\end{align}
The variational empirical estimation of $\mathcal{L}_k^{\text{C-DMIB}}$ is given by
\begin{align}
\label{C-VDMIB-emp}
\mathcal{L}_k^{\text{C-VDMIB}}\simeq\frac{1}{B}\sum\nolimits_{i=1}^{B}[\mathcal{L}_k^S+\sum\nolimits_{m}\mathcal{L}_k^U],
\end{align}
where
\begin{align}
	\label{L_K_S}
	\!\!\!\!\!\mathcal{L}_k^S\!\!=&\zeta D_\text{KL}\left({\mathcal{N}(\boldsymbol{\mu}_k^{i}, \boldsymbol{\Sigma}_k^{i})||\mathcal{N}(\mathbf{0},\mathbf{I})}\right) 
	 -\log q_{\boldsymbol{ \theta}_B}(\boldsymbol{y}_k^i|\boldsymbol{ s}_k^{i}),\!\! \!\! \\	\label{L_K_U}
	 \!\!\!\!\!\mathcal{L}_k^U\!\!= &\zeta D_\text{KL}\big({\mathcal{N}(\boldsymbol{\mu}_k^{m,i}, \boldsymbol{\Sigma}_k^{m,i})||\mathcal{N}(\mathbf{0},\mathbf{I})}\big) 
	  \!\!- \!\! \log q_{\boldsymbol{ \theta}_B}(\boldsymbol{y}_k^i|\boldsymbol{ z}_k^{m,i}).\!\!\!\!
\end{align}
Particularly, the mean absolute error and cross-entropy loss are commonly employed to minimize the conditional entropy between the target and the extracted semantic representation for regression and classification tasks, respectively\cite{CL-Semcom}.


\subsection{Adaptive Semantic Feature Transmission}
Although the chosen variational approximation encourages sparsity in the transmitted semantic features to reduce communication overhead, an effective approach is still required to identify which transmission dimensions can be selected to be pruned.
More importantly, transmitting a larger number of features can enhance robustness against channel noise while increasing communication overhead.
Therefore, we propose a novel adaptive semantic feature transmission method
to cope with dynamic channel conditions.
Given a fully-connected (FC) layer, the computation operation is defined as 
\begin{align}
	\label{FC_layer}
\mathrm{FC}(\boldsymbol{a})=\boldsymbol{W}\boldsymbol{a}+\boldsymbol{b},
\end{align}
where $\boldsymbol{a}$, $\boldsymbol{b}$, and $\boldsymbol{W}$ indicate the input, bias, and weight matrix, respectively.
Define $\boldsymbol{\bar W}=[\boldsymbol{W},\boldsymbol{b}]$ as an augmented weight matrix with $\boldsymbol{\bar W}_j$ being the $j$-th row of $\boldsymbol{\bar W}$.
Let $\phi_j$ denote the $j$-th dimension of $\boldsymbol\phi$, which is the scaling factor for each row in $\boldsymbol{\bar W}$.
Then, the proposed adaptive transmission approach establishes the mapping from the input $\boldsymbol x_k^m$ to the encoded semantic feature $\boldsymbol z_k^m$ based on the following formula:
\begin{align}
	\label{adaptive_feature _trans}
	\boldsymbol z_{k,j}^m= \operatorname{Tanh}\left(S^{\mathcal{T}}(\boldsymbol{x}_k^m)\frac{\boldsymbol{\bar W}_j}{\|\boldsymbol{\bar W}_j\|_2}\phi_j(\sigma^2)\right),
\end{align}
where $S^{\mathcal{T}}(\cdot)$ is the function of the semantic encoder, $\operatorname{Tanh}(\cdot)$ denotes the activation function, $\boldsymbol z_{k,j}^m$ is the $j$-th dimension of $\boldsymbol z_{k}^m$,
and $\phi_j(\sigma^2)$ (the $j$-th item of $\boldsymbol \phi(\sigma^2)$) is the semantic feature importance determined by the channel noise variance $\sigma^2$.
Under poor channel conditions, we obtain a large $\phi_j(\sigma^2)$ and $\boldsymbol z_{k,j}^m$ approaches 1, 
necessitating the transmission of a large number of semantic features to mitigate the impact of channel noise.
Conversely, when channel conditions are favorable, transmitting a smaller number of semantic features is sufficient.
More importantly, to further reduce the transmission overhead,
the activated dimensions are sequentially assigned starting from the first dimension, eliminating the need for additional communication resources to transmit their indices or mask elements like in the existing works\cite{Unified-SemCom,SemCom-Memory}.
The improved semantic dimension importance $\phi_j(\sigma^2)$ is given by
\begin{align}
\label{semantic_importance}
\phi_j(\sigma^2)=\sum\nolimits_{r=j}^d \psi_r(\sigma^2),
\end{align}
where $d$ is the dimension of the encoded semantic vector, and $\psi_r(\cdot)$ is the $r$-th output dimension of the function $\boldsymbol\psi(\cdot)$ parameterized by a 
multi-layer perceptron (MLP).
By restricting the parameter range of MLP, $\psi_r(\sigma^2)$ is ensured to be a non-negative increasing function, naturally leading to $\phi_j(\sigma^2)\leq\phi_r(\sigma^2),\forall j > r$ and $\phi_j(\sigma_1^2)\leq\phi_j(\sigma_2^2),\forall \sigma_1^2 \leq \sigma_2^2,$
Hence, given a threshold $\phi_{\text{th}}$, the proposed adaptive transmission method can flexibly adjust the dimensions of the transmitted semantic features in a successive manner under dynamic channel conditions.

\subsection{Task-Agnostic Semantic Transceiver Structure}
As shown in Fig. \ref{TASC-framework}, each semantic transmitter is comprised of a task-oriented semantic encoder and a dynamic channel encoder, and the semantic receiver includes a task-agnostic channel decoder and a task-agnostic semantic decoder.

\subsubsection{Task-Oriented Semantic Transmitter}
Thanks to the powerful semantic modeling and representation extraction capabilities,
we primarily employ the Transformer\cite{Transformer} encoder as the backbone network to construct the semantic encoder for visual, linguistic, and acoustic modalities.
Specifically, each Transformer encoder layer is composed of two components: self-attention network (SAN) and feed-forward network (FFN).
The core is the attention mechanism of SAN that is calculated 
by adopting the dot-product similarity function.
To clarify, SAN first projects the input sequence $\boldsymbol X_{\text{in}}^m$ of the modality $m$ to \emph{query}, \emph{key}, and \emph{value} space, respectively, i.e.,
\begin{align}
	\label{QKV}
	\boldsymbol Q= \boldsymbol{W}_q {\boldsymbol X_{\text{in}}^m}, \quad \boldsymbol{K}= \boldsymbol{W}_k {\boldsymbol X_{\text{in}}^m}, \quad {\boldsymbol V}=\boldsymbol{W}_v {\boldsymbol X_{\text{in}}^m},
\end{align}
where $\boldsymbol {W}_q$, $\boldsymbol {W}_k$, and $\boldsymbol {W}_v$ are learnable parameters. 
Then, the output of self-attention can be attained by
\begin{align}
	\label{SA}
	\mathrm{SA}(\boldsymbol X_{\text{in}}^m)=\operatorname{Softmax}({\boldsymbol Q^{\top} \boldsymbol{K}}/{\sqrt{N_d}} )\boldsymbol{V}.
\end{align}
where $N_d$ is the hidden size.
In SAN, multi-head self attention (MSA) is adopted to promote the capacity, where multiple parallel attention heads are employed individually on the input sequence. The output of each head is concatenated and linearly transformed into the final output. 
Thus we have
\begin{align}	
\nonumber
&\!\!\! \mathrm{MSA}(\boldsymbol X_{\text{in}}^m)= \\   \label{MHA}
&\!\!\! \boldsymbol X_{\text{in}}^m+(\mathrm{SA}(\boldsymbol X_{\text{in}}^m)\oplus\mathrm{SA}(\boldsymbol X_{\text{in}}^m)\oplus\cdots\oplus \mathrm{SA}(\boldsymbol X_{\text{in}}^m))\boldsymbol W_{H},
\end{align}
where $\boldsymbol W_{H}$ is learnable parameters.
Afterwards, $\mathrm{MSA}(\boldsymbol X_{\text{in}}^m)$ is fed into FFN, and it can be obtained that
\begin{align}	
\label{Transformer}
\boldsymbol X_{\text{out}}^m=\mathrm{MSA}(\boldsymbol X_{\text{in}}^m)+\mathrm{FFN}(\mathrm{MSA}(\boldsymbol X_{\text{in}}^m)).
\end{align}
To handle multiple tasks with a single set of parameters,
the encoder of TASC must identify the current task to perform targeted feature extraction effectively.
Therefore, the task embedding vector $\boldsymbol{\varpi}_k^m$ is introduced and trained across the whole network.
By concatenating $\boldsymbol X_{\text{in}}^m$, the input of the semantic encoder layer is  denoted as
\begin{align}	
	\label{task embedding vec}
	\boldsymbol {\widetilde{X}}_{\text{in}}^m=\boldsymbol X_{\text{in}}^m\oplus\boldsymbol{\varpi}_k^m.
\end{align}
To facilitate the transmission and resist the dynamic channel conditions, an adaptive channel encoder composed of multiple MLP layers is developed for flexible semantic feature transmission w.r.t. ${\boldsymbol{Z}_k^m}$ as discussed in Section II.C. 

\subsubsection{Task-Agnostic Semantic Receiver}
At the semantic receiver, the received noise-corrupted features ${\boldsymbol{\hat Z}_k^m}$ are firstly fed into a task-agnostic channel decoder, consisting of multiple MLP layers, to mitigate the impact of channel noise.
Subsequently, the task-agnostic semantic decoder processes the output of the channel decoder to generate task-specific multimodal representations, which are then utilized for inference.
The task-agnostic semantic decoder is built upon the unified Transformer\cite{Unified-Transformer} decoder structure, which is comprised of three  components: SAN, cross-attention network (CAN), and FFN.
Unlike SAN, CAN is designed to learn essential information among various modalities. 
Given the input $\boldsymbol{Z}_{\text{in}}^1$ and $\boldsymbol{Z}_{\text{in}}^2$, it can be obtained that
\begin{align}
	\label{QKV_CA}
	\boldsymbol Q_1= \boldsymbol{W}_q {\boldsymbol{Z}_{\text{in}}^1}, \quad \boldsymbol{K}_2= \boldsymbol{W}_k {\boldsymbol{Z}_{\text{in}}^2}, \quad {\boldsymbol V}_2=\boldsymbol{W}_v {\boldsymbol{Z}_{\text{in}}^2}.
\end{align}
Then the operation of CAN is denoted as 
\begin{align}
	\label{CAN}
	\mathrm{CA}(\boldsymbol{Z}_{\text{in}}^1,\boldsymbol{Z}_{\text{in}}^2)=\operatorname{Softmax}(\beta \boldsymbol Q_1^{\top} \boldsymbol{K}_2)\boldsymbol{V}_2.
\end{align}
Similar to MSA, multi-head cross attention (MCA) performs
\begin{align}	
	\nonumber
	 \mathrm{MCA}(\boldsymbol{Z}_{\text{in}}^1,\boldsymbol{Z}_{\text{in}}^2)=& \boldsymbol{Z}_{\text{in}}^1+(\mathrm{CA}(\boldsymbol{Z}_{\text{in}}^1,\boldsymbol{Z}_{\text{in}}^2)\oplus\mathrm{CA}(\boldsymbol{Z}_{\text{in}}^1,\boldsymbol{Z}_{\text{in}}^2)\\   \label{MCA}
	& \oplus\cdots\oplus \mathrm{CA}(\boldsymbol{Z}_{\text{in}}^1,\boldsymbol{Z}_{\text{in}}^2))\boldsymbol W_{H}.
\end{align}
The output of a Transformer decoder layer is given by
\begin{align}	
	\label{U-Transformer}
	\boldsymbol{Z}_{\text{out}}=\mathrm{MSA}(\boldsymbol{Z}_{\text{in}}^1)+\mathrm{FFN}(\mathrm{MCA}(\mathrm{MSA}(\boldsymbol{Z}_{\text{in}}^1),\boldsymbol{Z}_{\text{in}}^2)).
\end{align}

Specifically, the decoded features by the channel decoder are first processed by the fusion network $\boldsymbol Z_k={\boldsymbol{\hat Z}_k^v}\oplus{\boldsymbol{\hat Z}_k^l}\oplus{\boldsymbol{\hat Z}_k^a}$. 
Then, a semantic filter network including multiple Unified Transformer decoder processes $\boldsymbol Z_k$ to filter out redundant task-irrelevant information with semantic noise caused by multimodal feature fusion, and produces $\boldsymbol S_k$.
Finally, $\boldsymbol S_k$ are directly fed into the universal decoder network that consists of multiple Unified Transformer decoder layers to infer the final prediction.
Notably, the first Transformer decoder layer takes $\boldsymbol Z_k$ and learnable task-specific query embedding matrix $\boldsymbol T_K$ as the input, i.e., $\boldsymbol{Z}_{\text{in}}^1=\boldsymbol T_K$ and $\boldsymbol{Z}_{\text{in}}^2=\boldsymbol Z_k$ in (\ref{U-Transformer}), given by
\begin{align}	
	\label{First-decoder}
	\boldsymbol{Z}_{\text{out}}^1=\mathrm{MSA}(\boldsymbol T_K)+\mathrm{FFN}(\mathrm{MCA}(\mathrm{MSA}(\boldsymbol T_K),\boldsymbol{Z}_k)).
\end{align}
The task-specific query matrix $\boldsymbol T_K$ functions as an identifier for the task assigned to the semantic decoder, analogous to the conventional input utilized by the Transformer decoder.
For the subsequent decoder layers, the output of the $i$-th decoder layer $\boldsymbol{Z}_{\text{out}}^i$ can be iteratively represented as 
\begin{align}	
	\label{latent-decoder}
\!\! \boldsymbol{Z}_{\text{out}}^i = \mathrm{MSA}(\boldsymbol{Z}_{\text{out}}^{i-1})+\mathrm{FFN}(\mathrm{MCA}(\mathrm{MSA}(\boldsymbol{Z}_{\text{out}}^{i-1}),\boldsymbol{Z}_k)).
\end{align}

 \section{TASC Training Based on FML}
In this section, we present a novel distributed learning framework for TASC by exploiting the advantageous FML.

\subsection{FML Problem}

Our goal is to collaboratively meta-train the TASC system based on data distributed among SUs, such that the models can perform well for brand-new users and data via a few quick gradient descent steps.
For simplicity, we omit the modality $m$ and redefine the loss function of the $k$-th SU w.r.t. the parameter $\boldsymbol{\theta}$ as $l_{k}\left(\boldsymbol{\theta}\right)$.
In FML, each local dataset $\mathcal D_k$ is divided into the \emph{support} set $\mathcal D_k^{S}$ and the \emph{query} set $\mathcal D_k^{Q}$, both of which contain labeled data samples.
The objective of FML is 
\begin{eqnarray} 
	\label{vanilla-FML}
	\min _{\boldsymbol{\theta}}  L\left(\boldsymbol{\theta}\right)=  \sum\nolimits_{k \in \mathcal{K}} \omega_k L_k(\boldsymbol{\theta}_k)=\sum\nolimits_{k \in \mathcal{K}} \omega_k l_k(\boldsymbol\varphi_k,\mathcal D_k^{Q}),\!\!
\end{eqnarray}
where $\omega_k={D_k}/{\sum_{k \in \mathcal{K}}D_k}$ is the weight.
Concretely, 
at the beginning of communication round $t$, the BS selects a set $\mathcal K$ of $K$ SUs and distributes $\boldsymbol{\theta}(t)$ to them. 
For the $\tau$-th meta-learning step where $0 \leq \tau \leq \tau_0-1$,
the $k$-th user first performs \emph{inner update} using SGD based on $\mathcal D_k^{S}$, i.e.,
\begin{eqnarray} 
	\label{inner update}
\boldsymbol\varphi_k(t,\tau)=\boldsymbol{\theta}_k(t,\tau)-\alpha \nabla_{\boldsymbol{\theta}} l_k\left(\boldsymbol{\theta}_k(t,\tau),\mathcal D_k^{S}\right),
\end{eqnarray}
where $\alpha$ is the inner learning rate 
Then the meta-function $L_k(\boldsymbol{\theta}_k)=l_k(\boldsymbol{\varphi_k},\mathcal D_k^{Q})$ is evaluated based on $\boldsymbol\varphi_k$ and $\mathcal D_k^{Q}$ to perform \emph{outer update}, given by
\begin{eqnarray} 
	\label{outer update}
	\boldsymbol{\theta}_{k}(t,\tau+1)=\boldsymbol{\theta}_{k}(t,\tau)-\beta \tilde{\nabla} L_k\left(\boldsymbol\theta_k(t, \tau)\right),
\end{eqnarray}
where $\beta>0$ is the meta-learning rate. Note that $\boldsymbol{\theta}(t)=\boldsymbol{\theta}_{k}(t,0)$ when $\tau=0$ at the start, and $\tilde{\nabla} L_k\left(\boldsymbol\theta\right)$ is a biased estimate of $\nabla L_k\left(\boldsymbol\theta\right)$, expressed as
\begin{eqnarray} 
	\label{estimate} 
	\!\tilde{\nabla} \!L_k\!\big(\boldsymbol\theta\big)\!=\!\big(I\!-\!\alpha \tilde{\nabla}^2 l_k\big(\boldsymbol\theta; \mathcal{D}_k^{\prime\prime}\big)\big) \!\tilde{\nabla} l_k \!
	\big(\boldsymbol\theta\!-\!\alpha \tilde{\nabla} l_k\big(\boldsymbol\theta ; \mathcal{D}_k\big) ; \mathcal{D}_k^{\prime}\big)\!,\!
\end{eqnarray}
where $\mathcal{D}_k$, $\mathcal{D}_k^{\prime}$, and $\mathcal{D}_k^{\prime\prime}$ are independent batches, $\tilde{\nabla} l_k\left(\boldsymbol\theta\right)$ and $\tilde{\nabla}^2 l_k\left(\boldsymbol\theta\right)$ are the unbiased estimates
of $\nabla l_k\left(\boldsymbol\theta\right)$ and ${\nabla}^2 l_k\left(\boldsymbol\theta\right)$, respectively.
After running $\tau_0$ meta-learning
steps, each user transmits its meta model $\boldsymbol{\theta}_{k}(t)=\boldsymbol{\theta}_{k}(t,\tau_0)$ to the BS.
Then the BS updates the global model by aggregating the received models to obtain ${\boldsymbol{\theta}}(t+1)$, which is given by
\begin{eqnarray} 
{\boldsymbol{\theta}}(t+1) =  \sum\nolimits_{k \in \mathcal{K}} \omega_k \boldsymbol{\theta}_{k}(t).
\end{eqnarray}

It is worth noting that FL can be deemed as a special case of FML when $\alpha$ in (\ref{inner update}) is set to 0.
On the one hand, FML provides a personalized solution that effectively
accounts for the heterogeneity among SUs.
On the other hand, 
instead of locally updating the model directly as in FL, FML targets to learn a well-initialized model $\boldsymbol{\theta}$ which can quickly adapt to new users and data with only a small number of samples. 



\subsection{TASC Training Framework}

A reasonable multiuser SemCom model is inherently partitioned into multiple semantic transmitters at the SUs and a shared receiver network at the BS.
Inspired by this, we present our novel distributed learning architecture for TASC by leveraging FML.
The detailed learning procedure is outlined in Algorithm \ref{TASC algorithm}.
For each training round $t\in\mathcal{T}=\{0,1,\ldots,C-1\}$, the process consists of the following inner learning, outer learning, and federated aggregation stages.

\subsubsection{Inner Learning Stage}
The first inner learning stage of TASC is to facilitate the inner update for multiuser SermCom systems, involving the following steps.

\begin{itemize}

\item \emph{SU-side Inner Model Forward Propagation (SIMFP):}
In this step, all participating SUs perform the inner FP for semantic transmitters in parallel. Specifically, a mini-batch data with $b$ samples is randomly drawn from SU $k$, indicated by $\mathcal{B}_{k}^{I}(t,\tau)\subseteq \mathcal{D}_{k}^{S}$.
Let $\mathbf{X}_{k}^{I}(t,\tau)\in\mathbb{R}^{b \times q}$ and $\mathbf{y}_{k}(t,\tau)\in\mathbb{R}^{b \times 1}$ denote the aggregated input data and labels of the mini-batch, respectively.
Then the input data are fed into the SU-side transmitter network to produce transmitted signals, denoted as $\mathbf{Z}_{k}^{I}(t,\tau)\in\mathbb{C}^{b \times c}$, i.e.,
\begin{align}
	\label{inner FP SU}
\mathbf{Z}_{k}^{I}(t,\tau)=\mathcal {ST}(\mathbf{X}_{k}^{I}(t,\tau);\boldsymbol{\theta}_{k}(t,\tau)).
\end{align}

\item \emph{Inner Semantic Data Transmission (ISDT):}
After completing the inner FP process for the SU-side transmitter network, 
each participating user transmits signals and labels to the BS.
The received signal is then utilized as the input of the BS-side receiver network to accomplish the remaining part of the inner FP process.

\item  \emph{BS-side Inner Model Forward and Backward Propagation (BIMFBP):}
Subsequently, the BS collects the transmitted signals that are concatenated into matrix $\mathbf{\hat Z}^{I}(t,\tau)=[\mathbf{\hat Z}_{1}^{I}(t,\tau);\mathbf{\hat Z}_{2}^{I}(t,\tau);\cdots;\mathbf{\hat Z}_{K}^{I}(t,\tau)]\in\mathbb{C}^{Kb \times c}$, which is then fed into the receiver network $\boldsymbol{\theta}_{B}(t,\tau)$.
Thus, the predicted result is obtained by
\begin{align}
\label{inner FP BS}
\mathbf{\hat{y}}^{I}(t,\tau)=\mathcal {SR}(\mathbf{\hat Z}^{I}(t,\tau);\boldsymbol{\theta}_B(t,\tau)).
\end{align}
With the knowledge of (\ref{inner FP SU}) and (\ref{inner FP BS}), the one-round FP process of the inner update is finished.
Given the predicted results and the corresponding labels, the average gradients of the loss function can be computed to perform the BP process of the inner update via SGD:
\begin{align}
	\label{inner BP BS}
	\boldsymbol{\varphi}_B(t,\tau)=\boldsymbol{\theta}_B(t,\tau)-\alpha_B \nabla_{\boldsymbol{\theta}} l_k(\boldsymbol{\theta}_B(t,\tau)),
\end{align}
where $\boldsymbol{\varphi}_B(t,\tau)$ indicates the parameters of the BS-side model for inner update and $\alpha_B$ is the inner learning rate.

\item \emph{Inner Activations' Gradients Transmission (IAGT):}
When the inner BP process reaches the first layer of the receiver network, 
the activations' gradients of a mini-batch data, 
are transmitted to the corresponding SUs.

\item \emph{SU-side Inner Model Backward Propagation (SIMBP):}
Then each SU-side transmitter model is updated via SGD:
\begin{align}
	\label{inner BP SU}
	\boldsymbol{\varphi}_{k}(t,\tau)=\boldsymbol{\theta}_{k}(t,\tau)-\alpha_S \nabla_{\boldsymbol{\theta}} l_k(\boldsymbol{\theta}_{k}(t,\tau)),
\end{align}
where $\boldsymbol{\varphi}_{k}(t,\tau)$ is the parameters of the SU-side model after inner update, and $\alpha_S$ is the inner learning rate.

\end{itemize}

 \subsubsection{Outer Learning Stage}
Similar to the inner learning process, this stage is to evaluate parameter $\boldsymbol{\varphi}$ on the query set
by computing the test loss, which reflects the training ability of the semantic model. Then, the model parameters are further updated by minimizing the test loss, i.e., outer update.

\begin{itemize}

\item \emph{SU-side Outer Model Forward Propagation (SOMFP):}
Given the input data $\mathbf{X}_{k}^{O}(t,\tau)$ and parameter $\boldsymbol{\varphi}_{k}(t,\tau))$, the FP process of the outer update is performed as 
\begin{align}
	\label{outer FP SU}
	\mathbf{Z}_{k}^{O}(t,\tau)=\mathcal {ST}(\mathbf{X}_{k}^{O}(t,\tau);\boldsymbol{\varphi}_{k}(t,\tau)).
\end{align}

\item \emph{Outer Semantic Data Transmission (OSDT):}
Then the produced transmitted signals and labels are sent to the BS to complete the FP process of the outer update.

\item \emph{BS-side Outer Model Forward and Backward Propagation (BOMFBP):}
With the received signals, the BS concatenates them into
matrix $\mathbf{\hat Z}^{O}(t,\tau)\in\mathbb{C}^{Kb \times c}$.
Then, the predicted value is obtained by
\begin{align}
\label{outer FP BS}
\mathbf{\hat{y}}^{O}(t,\tau)=\mathcal {SR}(\mathbf{\hat Z}^{O}(t,\tau);\boldsymbol{\varphi}_B(t,\tau)).
\end{align}
Note that the outer learning stage directly updates $\boldsymbol{\theta}(t,\tau)$ rather than $\boldsymbol{\varphi}(t,\tau)$ via SGD, given by
\begin{align}
\label{outer BP BS}
\boldsymbol{\theta}_B(t,\tau+1)=\boldsymbol{\theta}_B(t,\tau)-\beta_B \nabla_{\boldsymbol{\theta}} l_k(\boldsymbol{\varphi}_B(t,\tau)),
\end{align}
where $\beta_B$ denotes the meta-learning rate at the BS.

\item \emph{Outer Activations' Gradients Transmission (OAGT):}
After completing the BS-side outer BP process, the activations' gradients are sent to SUs for the BP process.

\item \emph{SU-side Outer Model Backward Propagation (SOMBP):}
In this step, each SU performs the meta-update for the semantic transmitter network based on SGD, given by
\begin{align}
\label{outer BP SU}
\boldsymbol{\theta}_{k}(t,\tau+1)=\boldsymbol{\theta}_{k}(t,\tau)-\beta_S \nabla_{\boldsymbol{\theta}} l_k(\boldsymbol{\varphi}_{k}(t,\tau)),
\end{align}
where $\beta_S$ is the SU's meta-learning rate.

\end{itemize}

\subsubsection{Federated Aggregation Stage}
After completing $\tau_0$ steps of inner and outer learning, SU-side models are uploaded and aggregated based on the FedAvg algorithm, expressed by
\begin{align}
\label{aggregation}
{\boldsymbol{\theta}}(t+1) =  \frac{\sum_{k \in \mathcal{K}} D_k \boldsymbol{\theta}_{k}(t, \tau_0)}{\sum_{k \in \mathcal{K}}D_k}.
\end{align}
Note that $\boldsymbol{\theta}_{k}(t+1,0)={\boldsymbol{\theta}}(t+1)$
when $\tau=0$ at the beginning of the $(t+1)$-th communication round.
%

\begin{algorithm}[htbp]
	\caption{Training Process of The TASC Framework}
	\label{TASC algorithm}
	\textbf{Input:} $\ \!\!\! \  b, \alpha_B, \alpha_S, \beta_B,  \beta_S,  \tau_0, C$;
	\begin{algorithmic}[1]
		\STATE Initialize model parameters $\boldsymbol\theta_k$ and $\boldsymbol\theta_{B}$;
		\WHILE{communication round $t=0$ to $C-1$}
		\STATE BS broadcasts the latest transmitter model to SUs;
		\FOR {meta-learning step $\tau=0$ to $\tau_0-1$ }
		\FOR {each SU $k\in\mathcal{K}$ in parallel}
		\STATE Draw a mini-batch of data samples;
		\STATE Execute SU-side inner model FP based on (\ref{inner FP SU});
		\STATE Send semantic data to the BS;
		\ENDFOR
		\STATE BS performs model FP and BP via (\ref{inner FP BS}) and (\ref{inner BP BS});
		\STATE BS transmits activations' gradients to SUs;
		\FOR {each SU $k\in\mathcal{K}$ in parallel}
		\STATE Update the semantic transmitter model via (\ref{inner BP SU});
		\STATE Execute SU-side outer model FP based on (\ref{outer FP SU});
		\STATE Transmit semantic data to the BS;
		\ENDFOR
		\STATE BS executes FP and BP based on (\ref{outer FP BS}) and (\ref{outer BP BS}); 
		\STATE BS sends the outer activations' gradients to SUs;
		\FOR {each SU $k\in\mathcal{K}$ in parallel}
		\STATE Update the semantic transmitter model via (\ref{outer BP SU});
		\ENDFOR
		\ENDFOR
		\FOR {each SU $k\in\mathcal{K}$ in parallel}
		\STATE Upload the SU-side transmitter models to the BS;	
		\STATE BS computes loss divergence $\delta_k(t)$ based on (\ref{delta});	
		\ENDFOR	
		\STATE BS determines SU selection $\mathcal{K}^t$ by solving (\ref{P-DQ})
		\STATE BS aggregates SU-side transmitter models via (\ref{aggregation});
		\ENDWHILE
	\end{algorithmic}
	\textbf{Output:} The converged semantic transceiver model $\boldsymbol{\theta}$.
\end{algorithm}

\section{Theoretical Analysis}

\subsection{Assumptions}
For a more general case, non-convex settings are considered\cite{FML-Personalized}.
The aim is to identify an $\epsilon$-approximate first-order stationary point (FOSP) in problem (\ref{vanilla-FML}), defined as follows.
\begin{definition}
\label{Definition 1}
\emph{
Define a random vector $\boldsymbol\theta_\epsilon\in\mathbb{R}^{d}$ as an $\epsilon$-FOSP in problem (\ref{vanilla-FML}) if it satisfies $\mathbb{E}[\left\|\nabla L\left(\boldsymbol\theta_\epsilon\right)\right\|^2] \leq \epsilon$.
}
\end{definition}

Then the assumptions are stated without loss of generality.
\begin{assumption}
	\label{Assumption 1}
	\emph{
For every SU $k\in\mathcal{K}$, the expected loss function $l_k$ is twice continuously differentiable and $H_k$-smooth, which is 
\begin{align}
	\label{assum1}
\left\|\nabla l_k(\boldsymbol\theta_1)-\nabla l_k(\boldsymbol\theta_2)\right\| \leq H_k\|\boldsymbol\theta_1-\boldsymbol\theta_2\|,  \forall \boldsymbol\theta_1,\boldsymbol\theta_2 \in \mathbb{R}^{d}.
\end{align}
Besides, its gradient is bounded by a positive constant $\Phi_k$, i.e., $\left\|\nabla l_k(\boldsymbol\theta)\right\|\leq\Phi_k$.
}
\end{assumption}

\begin{assumption}
\label{Assumption 2} 
\emph{
The Hessian of $l_k(\boldsymbol\theta)$ is $\lambda_k$-Lipschitz continuous, which is
\begin{align}
	\label{assum2}
\!\!\!\!	\left\|\nabla^2 l_k(\boldsymbol\theta_1)-\nabla^2 l_k(\boldsymbol\theta_2)\right\| \leq \lambda_k\|\boldsymbol\theta_1-\boldsymbol\theta_2\|,  \forall \boldsymbol\theta_1,\boldsymbol\theta_2 \in \mathbb{R}^{d}.
\end{align}
}
\end{assumption}

\begin{assumption}
\label{Assumption 3} 
\emph{
For any $\boldsymbol{\theta}\in\mathbb{R}^{d}$, the gradient $\nabla\ell_k(\boldsymbol{\theta}; \boldsymbol x, y)$ and Hessian $\nabla^2\ell_k(\boldsymbol{\theta}; \boldsymbol x, y)$ w.r.t. a single point $(\boldsymbol{x}_{k}^{i},y_{k}^{i})\in \mathcal{X} \times \mathcal{Y}$, 
satisfy the following conditions
	\begin{align}
		\label{assum3-1}
&	\mathbb{E}_{(x, y) \sim P_k}[||\nabla\ell_k(\boldsymbol{\theta}; \boldsymbol x, y)-\nabla l_k(\boldsymbol\theta)||^2]\leq \sigma_G^2, \\ 		\label{assum3-2}
& 	\mathbb{E}_{(x, y) \sim P_k}[||\nabla^2\ell_k(\boldsymbol{\theta}; \boldsymbol x, y)-\nabla^2 l_k(\boldsymbol\theta)||^2]\leq \sigma_H^2.
\end{align}
}
\end{assumption}

\begin{assumption}
\label{Assumption 4}
\emph{
For any $\boldsymbol{\theta}\in\mathbb{R}^{d}$, the following facts hold for gradient and Hessian of $l_k\left(\boldsymbol\theta\right)$ and $l\left(\boldsymbol\theta\right)$
	\begin{align}
	\label{assum4-1}
	&\frac{1}{K} \sum\nolimits_{k=1}^K\left\|\nabla l_k(\boldsymbol{\theta})-\nabla l(\boldsymbol{\theta})\right\|^2 \leq \mu_G^2, \\ 		\label{assum4-2}
	& 	\frac{1}{K} \sum\nolimits_{k=1}^K\left\|\nabla^2 l_k(\boldsymbol{\theta})-\nabla^2 l(\boldsymbol{\theta})\right\|^2 \leq \mu_H^2,
\end{align}
where $l\left(\boldsymbol\theta\right)=\sum\nolimits_{k=1}^Kl_k\left(\boldsymbol\theta\right)$ is the average function of $l_k\left(\boldsymbol\theta\right)$.
}
\end{assumption}

\subsection{Convergence Analysis}
Denote $\Phi=\max_k\Phi_k$, $H=\max_kH_k$, $\lambda=\max_k\lambda_k$ for ease of analysis.
Then the following lemmas are provided.

\begin{lemma}
	\label{Lemma 1}
\emph{
	If Assumptions \ref{Assumption 1} and \ref{Assumption 2} hold, then for any $\alpha\in(0,1/H]$, the meta-function of SU $k$, i.e., $L_k(\boldsymbol{\theta})$, is smooth with parameter $H_L=4H+\alpha\lambda\Phi$, and the average function $L(\boldsymbol{\theta})$ is also smooth with $H_L$.
}
\end{lemma}

\begin{lemma}
	\label{Lemma 2}
	\emph{
	If Assumptions \ref{Assumption 1}, \ref{Assumption 2}, and \ref{Assumption 3} hold, for any $\alpha\in(0,1/H]$ and $\boldsymbol{\theta}\in\mathbb{R}^{d}$,
	the following holds
	\begin{align}
		\label{lemma2-1}
		&\big\|\mathbb{E}\big[\tilde{\nabla}L_k(\boldsymbol{\theta})-\nabla L_k(\boldsymbol{\theta})\big]\big\| \leq {2 \alpha H \sigma_G}/{\sqrt{D}_k}, \\ 		\label{lemma2-2}
		& \mathbb{E}\big[{\big\|\tilde{\nabla}L_k(\boldsymbol{\theta})-\nabla L_k(\boldsymbol{\theta})\big\|}^2\big] \leq \sigma_L^2,
	\end{align}
	where $\sigma_L^2$ is denoted as
	\begin{align}
		\sigma_L^2= 12\big(\Phi^2+\sigma_G^2\big(\frac{D_k+D_k^{\prime}}{D_kD_k^{\prime}}\big)\big)\big(1+\sigma_H^2 \frac{\alpha^2}{4 D_k^{\prime \prime}}\big).
	\end{align}
}
\end{lemma}


\begin{lemma}
	\label{Lemma 3}
	\emph{
	Suppose that conditions in Assumptions \ref{Assumption 1}, \ref{Assumption 2}, and \ref{Assumption 4} are satisfied, for any $\boldsymbol{\theta}\in\mathbb{R}^{d}$, we have
	\begin{align}
		\frac{1}{K} \sum\nolimits_{k=1}^K\left\|\nabla L_k(\boldsymbol{\theta})-\nabla L(\boldsymbol{\theta})\right\|^2 \leq \mu_L^2,
	\end{align}
	where  $\mu_L^2$ is defined as
	\begin{align}
		\mu_L^2=3 \Phi^2 \alpha^2 \mu_H^2+192 \mu_G^2.
	\end{align}
}
\end{lemma}


With Lemmas \ref{Lemma 1}-\ref{Lemma 3},
the expected convergence result of TASC framework is obtained below.

\begin{theorem}
\label{theorem 1}
\emph{
Suppose that Assumptions \ref{Assumption 1}-\ref{Assumption 4} hold. Given independent sampled datasets with ${D}_k={D}_k^{\prime}={D}_k^{\prime\prime}=D$, the number of communication rounds $K$, and the optimal global loss $L\left(\boldsymbol{\theta}_\epsilon\right)$ , and then for $\alpha\in(0,1/H]$ and $\beta\in(0,1/{H_L}]$, the following FOSP condition holds
\begin{align}
\nonumber
\!\!&\frac{1}{\tau_0C} \sum\nolimits_{t=0}^{C-1}\sum\nolimits_{\tau=0}^{\tau_0-1} \mathbb{E}\!\left[\left\|\nabla L\left(\bar{\boldsymbol{\theta}}(t)\right)\right\|^2\right] \!\leq\! \frac{4\left(L\left(\boldsymbol{\theta}_0\right)\!-\!L\left(\boldsymbol{\theta}_\epsilon\right)\right)}{\beta \tau_0C} \\ \label{FOSP}
\!\!&+ 4\beta H_L\left(2 \sigma_L^2+\frac{\mu_L^2}{(K-1)}(\frac{K}{K^t}-1)\right)+\frac{16 \alpha^2 H^2 \sigma_G^2}{D_k},
\end{align}
where $K^t={\sum\nolimits_{k=1}^{K}a_k}$ is the number of selected SUs with $a_k\in\{0,1\}$ being the binary SU selection variable, and $\bar{\boldsymbol{\theta}}(t)$ is the average of local updates $\boldsymbol{\theta}_k(t)$.
}
\end{theorem}

\begin{IEEEproof}
	See Appendix A of the technical report \cite{Appendix}.
\end{IEEEproof}

Theorem \ref{theorem 1} provides a lower bound on the gradient of global objective function $L$.
It shows that increases of communication rounds and meta-learning steps improve the convergence of the proposed TASC framework.
Furthermore, it implies that how the task similarity and different SU selection impact the convergence. While the overall convergence performance exhibits weak dependence on the total number of SUs, the convergence speed improves as the number of selected SUs increases.
Therefore, it is critical to quantify the loss contribution of each SU for facilitating the convergence.

\subsection{SU Convergence Contribution}
To improve the convergence, the main result is given below.
\begin{corollary}
\label{corollary 1}
\emph{
If Assumptions \ref{Assumption 1}-\ref{Assumption 4} hold, given independent sampled datasets with ${D}_k={D}_k^{\prime}={D}_k^{\prime\prime}$, and for $\alpha\in(0,1/H]$ and $\beta\in(0,1/{H_L}]$, we have
\begin{align}
\nonumber
		\!\!\!\!\! \mathbb{E}\left[L\left(\boldsymbol\theta(t+1)\right)\! -\!L\left(\boldsymbol\theta(t)\right)\right] \leq & \frac{\beta}{2} \mathbb{E}[\frac{ 1 } {K_t} \sum _ { k \in \mathcal { K } _ { t } } (  -\|\tilde{\nabla} L_k\left(\boldsymbol\theta_k(t)\right)\|^2   
		\\
		\label{user diver}
		&  \!\!\!\!\!\!\!\!  \!\!\!\!\!\!\!\! \!\!\!\!\!\!\!\!\!\!\!   + (\eta_1+{\eta_2}/{\sqrt{D_k}}) \|\tilde{\nabla} L_k\left(\boldsymbol\theta_k(t)\right)\|   )     ],
\end{align}
where $\eta_1$ and $\eta_2$ are positive constants, which are
\begin{align}
	\label{eta1}
	&\eta_1\geq  \sqrt{16\mu_G+4\alpha\Phi\mu_H}+\beta\sqrt{140(\mu_G^2+2\sigma_L^2)}, \\ \label{eta2}
	& \eta_2 \geq 24\sigma_G^2(4+\alpha^2\sigma_H^2)+6\alpha^2\Phi^2\sigma_H^2.
\end{align}
}
\end{corollary}

\begin{IEEEproof}
	See Appendix B of the technical report \cite{Appendix}.
\end{IEEEproof}

From Corollary \ref{corollary 1}, it suggests that a SU with a large gradient inherently accelerates the reduction of global meta-loss. However, a small size of the dataset hinders this course due to the large variance. Additionally, as SU dissimilarities increase, the upper bound in equation (\ref{user diver}) becomes less restrictive. Inspired by this, the $k$-th SU loss divergence for convergence contribution in the $t$-th communication round is 
\begin{align}
	\label{delta}
	\!\!\!\!\!\!\!\!\! \delta_k(t)\!=\!(\eta_1\!+\!{\eta_2}/{\sqrt{D_k}}) \|\tilde{\nabla} L_k\left(\theta_k(t)\right)\| \!
 \!	-\! \!\|\tilde{\nabla} L_k\left(\theta_k(t)\right)\|^2.\!\!\!\!\!\!\!\!
\end{align}


\section{Resource Allocation for TASC Over \\ Wireless Networks}
In this section, we consider the proposed TASC framework over multi-access wireless networks.
First, we present the wireless transmission and computing model of TASC, followed by the analysis of training latency and energy cost.
We then perform effective resource allocation that jointly optimizes SU selection, power control, and computing capabilities.

\subsection{Wireless Transmission and Computation Model}
\subsubsection{Wireless Transmission Model}
For uplink transmission, the orthogonal frequency division multiple access (OFDMA)
is considered, whereby each SU occupies one uplink resource block (RB) in a round to transmit semantic data and upload its meta model. The achievable rate of SU $k$ is given by
\begin{align}
\label{R_up}
\!\!\!R_k^{U}(\boldsymbol a_k, p_k)\!=\!\sum\nolimits_{n \in \mathcal{R}} a_{k, n} W^{U} \log _2\left(1\!\!+\!\! \frac{h_k p_k}{I_n\!+\!W^{U} N_0}\right),\!\!\!\!
\end{align}
where $\boldsymbol a_k=[a_{k, 1},\ldots,a_{k, R}]$ indicates an RB allocation vector with $R$ being the number of RBs, denoted by $\mathcal{R}=\{1,2,\ldots,R\}$; $W^{U}$ is the bandwidth of each RB; 
$p_k$ is the transmit power of SU $k$; 
$N_0$ is the power spectral density of the Gaussian noise; $I_n$ is the interference caused by users located in other service areas who are utilizing RB $n$.
Each SU can occupy at most one RB while each RB can be assigned to at most one SU, which is specified as
\begin{align}
\!\!\!  \sum\nolimits_{n \in \mathcal{R}} a_{k, n} \leq 1 ,~ \forall k \in \mathcal{K}, 
~  \sum\nolimits_{k \in \mathcal{K}} a_{k, n} \leq 1 ,~  \forall n \in \mathcal{R}.\!\!\!\!\!
\end{align}
The downlink rate achieved by the BS to transmit gradients and parameters of global meta model to SU $k$, is denoted as
\begin{align}	
\label{R_down}
R_k^{D}=W^{D} \log _2\left(1+\frac{h_k p_B}{I^{D}+W^{D} N_0}\right),
\end{align}
where $W^{D}$ and $p_B$ are the bandwidth and transmit power of the BS, respectively; $I^{D}$ denotes the inference caused by other BSs which are not involved in the FML algorithm.

\subsubsection{Computation Model}
Let $f_k$ be the CPU-cycle frequency of SU $k$.
Denote $\gamma_k$ as the computation workload (in FLOPs) for SU $k$ to execute and update the model with one sample.
With $D_k$ being the batch size, the computation workload required to run one-step local training is $\gamma_kD_k$. Thus, the energy consumption of SU $k$ in a global round
is given by
\begin{align}	
\label{E_cp}
E_k^{cp}\left(f_k\right)=\varsigma_k \tau \kappa_k \gamma_k D_k f_k^2,
\end{align}
where $\varsigma_k$ is the effective capacitance coefficient,
$\kappa_k$ is the computing intensity (i.e., the number of CPU cycles required to complete one float-point operation) 
Then, the computation time of SU $k$ during the
local update phase is calculated as
\begin{align}	
\label{T_cp}
T_k^{c p}\left(f_k\right)={\tau \kappa_k \gamma_k D_k}/{f_k}.
\end{align}
Note that we set $\tau=1$ in the following for simplicity.

\vspace{-2mm}
\subsection{Training Latency and Energy Consumption}
\begin{figure}[t]
	\includegraphics[width=0.49\textwidth]{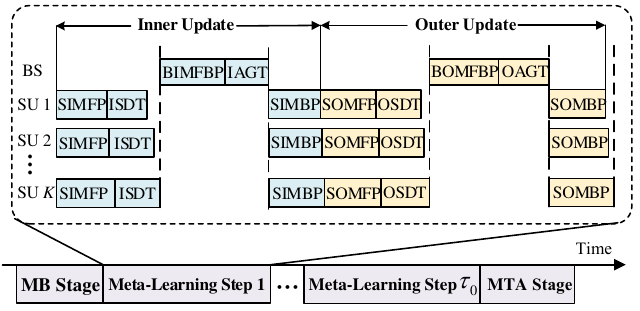} 
	\centering 
	\caption{The procedure of TASC in one communication round.}  
	\label{procedure}  
\end{figure}

Without loss of generality, we take into account one round analysis.
The index $m$ of modality is omitted for brevity.
As shown in Fig. \ref{procedure}, each round consists of four stages in a chronological manner: model broadcast, inner learning, outer learning, and model transmission and aggregation.
Each round involves the \emph{communication} and \emph{computation} processes.
Given the wireless transmission and computation model, the training latency and energy consumption of TASC are analyzed below.
\subsubsection{Model Broadcast (MB) Stage}
Denote $\xi_M$ as the model size of the transmitter network. 
The communication latency and energy consumption of MB are given by
\begin{align}	
\label{T_MB}
T_{B}^{D}={\xi_M}/{R_k^{D}}, \quad E_{B}^{D}=T_{B}^{D}p_B.
\end{align}

\subsubsection{Inner Learning Stage}
The computational latency and energy cost of SIMFP and SIMBP for SU $k$ are expressed by
\begin{align}	
\label{T_SIMFBP}
T_{k}^{cp,I}={ \kappa_k \gamma_k^{I} D_k}/{f_k},\quad 	E_{k}^{cp,I}=\varsigma_k  \kappa_k \gamma_k^{I} D_k f_k^2.
\end{align}
The computational latency and energy cost of BIMFBP are 
\begin{align}	
	\label{T_BIMFBP}
	T_{B}^{cp,I}\!\!=\!\!{ \sum\nolimits_{k } \! \kappa_B \gamma_B^{I} D_k}/{f_B},  ~ E_{B}^{cp,I}\!\!=\!\!\sum\nolimits_{k }\! \varsigma_B  \kappa_B \gamma_B^{I} D_k f_B^2,
\end{align}
where $\gamma_B^{I}$ is the computation workload of BS-side model's FP and BP process w.r.t. one data sample, $f_B$ is the CPU-cycle frequency of the BS. 
For semantic data uplink transmission, the latency and energy cost of ISDT are
\begin{align}	
\label{T_ISDT}
T_{k}^{co,I}={\xi_d^I}/{R_k ^{U}},\quad  E_{k}^{co,I}=\sum\nolimits_{n \in \mathcal{R}} a_{k, n} T_k^{co,I}p_k,
\end{align}
where $\xi_d^I$ is the transmitted semantic data size.
For the process of IAGT, the transmission latency and energy cost are 
\begin{align}	
	\label{T_IAGT}
	T_{B}^{co,I}={\xi_g^I}/{R_k^{D}}, \quad	E_{B}^{co,I}=T_{B}^{co,I}p_B,
\end{align}
where $\xi_g^I$ indicates the data size of activations' gradients.

\subsubsection{Outer Learning Stage}
Since the model parameters have a fixed dimension, the model sizes of semantic transceivers can be considered to be constant. Thereby we have
$\gamma_k^{O}=\gamma_k^{I}, \gamma_B^{O}=\gamma_B^{I}, \xi_d^o=\xi_d^I, \xi_g^O=\xi_g^I$. Consequently, it can be found that $T_{k}^{cp,I}=T_{k}^{cp,O},T_{B}^{cp,O}=T_{B}^{cp,I},T_{k}^{co,O}=T_{k}^{co,I},T_{g}^{co,O}=T_{g}^{co,I}$. The same is true for energy consumption, 
and hence, we omit superscripts $I$ and $O$ for simplicity.

\subsubsection{Model Transmission and Aggregation Stage (MTA)}
Regarding the MTA latency, each SU uploads its SU-side transmitter model to the BS using the allocated radio spectrum. 
Note that the aggregation latency is negligible as the FedAvg algorithm has relatively low computational complexity. Then, the MTA latency and energy cost are given by
\begin{align}	
	\label{T_MTA}
	T_{k}^{U}={\xi_M}/{R_k ^{U}},\quad 	E_{k}^{U}=\sum\nolimits_{n \in \mathcal{R}} a_{k, n} T_k^{U}p_k.
\end{align}

\vspace{-4mm}
\subsection{Problem Formulation}
With the analysis of all stages in (\ref{T_MB})-(\ref{T_MTA}), the one-round energy consumption and training latency are denoted as
\begin{align}
\label{Total_E}
\!\!\!\! \! \! E(\boldsymbol{a}, \boldsymbol{p}, \boldsymbol f)&\! =\!\sum\nolimits_{k \in \mathcal{K}}\! \! \left(E_k^{cp}\! +\!E_k^{co}\! +\!E_k^{U}\right) \! +\!E_{B}^{D}\! \! +\!E_B^{cp}\! +\!E_{B}^{co},	\! \! \!  \\ \nonumber
\!\!\!\! \! \!  T(\boldsymbol{a}, \boldsymbol{p}, \boldsymbol{f})&=\max\nolimits _{k \in \mathcal{K}} \{T_k^{c p}+T_B^{c o}+T_B^{D}\}+T_B^{cp}+ \\ \label{Total_T}
&  \! \ \ \ \ \max _{k \in \mathcal{K}} \sum\nolimits_{n \in \mathcal{R}} a_{k, n} (T_k^{c o}+T_k^{U}).
\end{align}
Moreover, to improve the convergence speed, we also need to minimize the following total SU contribution divergence as
\begin{align}
	\label{P-DQ}
 \Psi(\boldsymbol{a})=\sum
 \nolimits_{k \in \mathcal{K}}\sum\nolimits_{n \in \mathcal{R}}a_{k,n}\delta_k.
\end{align}

To address the heterogeneous channel conditions and computing capability of SUs while enhancing model convergence, we formulate the following optimization problem: 
 \begin{subequations}
 	\label{P0}
 	\begin{eqnarray}
 		\label{P0-function}
 \!\!\!\!\!\!\! \mathcal{P}:		& \underset{\boldsymbol{a}, \boldsymbol{p},\boldsymbol{f}}{\min}  &\Psi(\boldsymbol{a})+ \rho_1 E(\boldsymbol{a}, \boldsymbol{p}, \boldsymbol f)+\rho_2 T(\boldsymbol{a}, \boldsymbol{p}, \boldsymbol{f})    \\
 		\label{p_k}
 		&\operatorname{s.t.} &  0 \leq p_k \leq p_k^{max}, ~\forall k\in\mathcal{K}, \\
 		\label{f_k}
 		&&  0 \leq f_k \leq f_k^{max}, ~\forall k\in\mathcal{K}, \\
 		\label{a_k,n}
 		&&{a}_{k,n} \in \{0,1\}, ~\forall k\in\mathcal{K}, n \in \mathcal{R}, \\
 		\label{a_k}
 		&& \sum\nolimits_{n \in \mathcal{R}} a_{k, n} \leq 1 ,\quad \forall k \in \mathcal{K}, \\
 		\label{a_n} 
 		&& \sum\nolimits_{k \in \mathcal{K}} a_{k, n} \leq 1 ,\quad \forall n \in \mathcal{R},	 
 	\end{eqnarray}
 \end{subequations}
where $\rho_1$ and $\rho_2$ denote the weight coefficients that controls the Pareto-optimal tradeoff among convergence, latency, and energy cost.
Notably, $\mathcal{P}$ is a non-convex mixed-integer nonlinear programming (MINLP) problem with a severe straggler issue, which is challenging to solve.
To overcome this, a joint optimization algorithm
is provided.

\vspace{-4mm}
\subsection{Algorithm Design}
Note that since the BS equipped with a server has sufficient transmission and computing power, the computation and communication overhead of the BS is typically negligible.
To solve problem $\mathcal{P}$, it can be decomposed into two sub-problems:
 \begin{subequations}
\label{SP1}
\begin{eqnarray}
\label{SP1-function}
\!\!\!\!\!\!\!\!\!\! \mathcal{SP}1: \!&\!\!\!\!\!\underset{\boldsymbol{f}}{\min}   \ \ \!\!\!Q_1(\boldsymbol{f})\!\!=&\!\!\!\!\! \rho_1 \!\!\sum_{k \in \mathcal{K}}\varsigma_k  \kappa_k \gamma_k D_k f_k^2\!\!+\!\!\rho_2 \underset{k\in\mathcal{K}}{\max}\frac{ \kappa_k \gamma_k D_k}{f_k}   	\\
		\label{p_k_sp1}
		&\!\!\!\!\!\!\!\!\!\!\!\!\!\!\!\!\!\!\!\!\!\!\!\!\!\operatorname{s.t.} &\!\!\!\!\!\!\!\!\!\!\!\!\!\!\!\!\!\!\!\!\!\!\!\!\! (\ref{f_k}) .
	\end{eqnarray}
\end{subequations}
\vspace{-6mm}
\begin{subequations}
\label{SP2}
\begin{eqnarray}
\nonumber
\!\!\!\!\!\!\!\!\!\!\!\!\!\!\!\!\!\!\!\!\mathcal{SP}2:	\ \ \  &\!\!\!\!\!\!\!\!\!\!\!\!\!\!\!\!\!\!\!\!\!\!\!\!\!\!\!\!\underset{\boldsymbol{a},\boldsymbol{p}}{\min} \ \   Q_2(\boldsymbol{a},\boldsymbol{p})=\sum
\limits_{k \in \mathcal{K}}\sum\limits_{n \in \mathcal{R}}a_{k,n}\delta_k  \\ \nonumber
\ \ &+ \rho_1 \sum\limits_{k \in \mathcal{K}}\sum\limits_{n \in \mathcal{R}}a_{k,n}\frac{(\xi_d+\xi_M)p_k}{W^{U} \log _2(1+\frac{h_k p_k}{I_n+W^{U} N_0})}  \\ 	\label{SP2-function}&+ \rho_2 \underset{k\in\mathcal{K}}{\max}\sum\limits_{n \in \mathcal{R}}a_{k,n}\frac{\xi_d+\xi_M}{W^{U} \log _2(1+\frac{h_k p_k}{I_n+W^{U} N_0})}  \\
\label{ap_k_sp2}
\operatorname{s.t.}\!\!\!\!\!\!\!\!\!\!\! &\!\!\!\!\!\!\!\!\!\!\!\!\!\! (\ref{p_k}),(\ref{a_k,n}), (\ref{a_k}),(\ref{a_n}).
\end{eqnarray}
\end{subequations}

\subsubsection{CPU Frequency Optimization}
Problem (\ref{SP1}) is determined by the straggler's latency.
Assume that SU $l$ is the straggler user. 
Then $\mathcal{SP}1$ can be expressed as  
 \begin{subequations}
	\label{SP1-1}
	\begin{eqnarray}
		\label{SP1-1-function}
		& \underset{\boldsymbol{f}}{\min}   & \rho_1 \sum_{k \in \mathcal{K}}\varsigma_k  \kappa_k \gamma_k D_k f_k^2  +\rho_2 \frac{ \kappa_l \gamma_l D_l}{f_l}  \\
		\label{f_k_sp1}
		&\operatorname{s.t.} &\frac{ \kappa_k \gamma_k D_k f_l}{\kappa_l \gamma_l D_l} \leq {f_k}, ~  (\ref{f_k}). 
	\end{eqnarray}
\end{subequations}
\begin{lemma}
	\label{Lemma 4}
	\emph{The optimal solution to problem (\ref{SP1-1}) is 
\begin{align}	
\label{So-SP1-1}
f_k^*= \begin{cases}\min \left\{\sqrt[3]{\frac{b_1}{b_2}}, \underset{k \in \mathcal{K}}{\min} \frac{\kappa_l \gamma_l D_l f_l^{\max }}{\kappa_k \gamma_k D_k}\right\}, & \text { if } k=l \\ \frac{\kappa_k \gamma_k D_k f_l^*}{\kappa_l \gamma_l D_l}, & \text { otherwise }\end{cases}
\end{align}
where SU $l$ is the straggler among all SUs operating at optimal CPU frequencies $\boldsymbol{f}^*$, and  $b_1$ and $b_2$ are constants which are given by
\begin{align}
	\label{b1b2}
b_1=\rho_2 \kappa_l \gamma_l D_l, 
 b_2=2\rho_1 \sum_{k \in \mathcal{K}} \frac{\varsigma_k\left(\kappa_k \gamma_k D_k\right)^3}{\left(\kappa_l \gamma_l D_l\right)^2}.
\end{align}
}
\end{lemma}
\begin{IEEEproof}
	See Appendix C of the technical report \cite{Appendix}.
\end{IEEEproof}

From Lemma \ref{Lemma 4}, it suggests that if the straggler SU can be identified, 
the optimal CPU-cycle frequencies for all SUs can be obtained in closed-form solutions. 
This insights leads to the following Theorem.
\begin{theorem}
	\label{Theorem 2}
\emph{The global optimal solution to $\mathcal{SP}1$ is 
\begin{align}
	\label{So-SP1}
 \boldsymbol{f}^*={\operatorname{argmin}}_{f \in \{{f}_l^*\}_{l\in\mathcal{K}}}Q_1(\boldsymbol{f}),
\end{align}
where ${f}_l^*$ indicates the optimal solution of problem (\ref{SP1-1}) given the straggler SU $l$.
}
\end{theorem}

With Theorem \ref{Theorem 2}, $\mathcal{SP}1$ can be efficiently addressed with a computational complexity of $O(n)$ by evaluating the objective values for various potential straggler users.

\subsubsection{Transmit Power and RB Allocation}
Since $\mathcal{SP}2$ is a non-convex MINLP problem which is intractable to solve,
we present an efficient iterative algorithm in the following.
Define the optimal solutions of $\mathcal{SP}2$ as $\boldsymbol{a}^*=\{a_{k,n}|k\in\mathcal{K},n \in \mathcal{R}\}$, $\mathcal{K}^*=\{k\in\mathcal{K}|\sum\nolimits_{n \in \mathcal{R}} a_{k, n}= 1\}$, and $\boldsymbol{p}^*=\{p_k^*|k\in\mathcal{K}\}$, respectively.
Then the auxiliary variable $\chi$ is introduced to denote the transmission time, expressed as
\begin{align}
	\label{TransDelay}
	\chi^*=\underset{k\in\mathcal{K}}{\max}\sum
	\limits_{n \in \mathcal{R}}a_{k,n}^*\frac{(\xi_d+\xi_M)}{W^{U} \log _2(1+\frac{h_k p_k^*}{I_n+W^{U} N_0})},
\end{align}
 where it satisfies
 \begin{align}
 	\label{TransDelay_hold}
 	\chi^*\geq\frac{(\xi_d+\xi_M)}{W^{U} \log _2(1+\frac{h_k p_k^*}{I_n+W^{U} N_0})}.
 \end{align}
Given $\boldsymbol{a}^*$ and $\chi^*$, problem (\ref{SP2}) can be transformed to
 \begin{subequations}
	\label{SP2-1}
	\begin{eqnarray}
		\label{SP2-1-function}
		& \underset{\boldsymbol{p}}{\min}   & \sum
		\limits_{n \in \mathcal{R}}a_{k,n}^*\frac{\rho_1(\xi_d+\xi_M)p_k}{W^{U} \log _2(1+\frac{h_k p_k^*}{I_n+W^{U} N_0})}  \\
		\label{chi_sp2}
		&\operatorname{s.t.} &\sum
		\limits_{n \in \mathcal{R}}a_{k,n}^*\frac{(\xi_d+\xi_M)}{W^{U} \log _2(1+\frac{h_k p_k^*}{I_n+W^{U} N_0})} \leq {\chi^*},  \\
		\label{p_sp2}
		&&  (\ref{p_k}).
	\end{eqnarray}
\end{subequations}

\begin{lemma}
	\label{Lemma 5}
	\emph{Given the transmission time $\chi^*$, the optimal solution to problem (\ref{SP2-1}) is given by
	\begin{align}	
		\label{So-SP2-1}
p_k^*=\left\{\begin{array}{lc}
	\frac{\left(I_{n_k^*}+W^{U} N_0\right)\left(2^{\frac{\xi_d+\xi_M}{W^{U} \chi^*}-1}\right)}{h_k}, & \text { if } k \in \mathcal{K}^* \\
	0, & \text { otherwise.}
\end{array}\right.
\end{align}}
\end{lemma}		
\begin{IEEEproof}
	See Appendix D of the technical report \cite{Appendix}.
\end{IEEEproof}

Lemma \ref{Lemma 5} implies that the optimal solution of transmission power is determined in closed-form using (\ref{So-SP2-1}) with the knowledge of the RB allocation and transmission time.
Moreover, it states that (\ref{So-SP2-1}) can yield the optimal transmission power for any $\boldsymbol{a}^*$ and $\chi^*$, provided that condition (\ref{TransDelay_hold}) is satisfied.
Given $\chi^*$, problem (\ref{SP2}) is transformed to
\begin{subequations}
	\label{SP2-2}
	\begin{eqnarray}
	\label{SP2-2-function}
&\!\!\!\! \!\!\!\!\underset{\boldsymbol{a}}{\min} &\sum
		\limits_{k,n}a_{k,n}\left(\delta_k+\rho_1 \frac{(\xi_d+\xi_M)p_k}{W^{U} \log _2(1+\frac{h_k p_k}{I_n+W^{U} N_0})}\right)   \\
		\label{ap_k_sp2-2}
&\!\!\!\!\!\!\!\!\operatorname{s.t.} & (\ref{ap_k_sp2}).
	\end{eqnarray}
\end{subequations}
Based on Lemma \ref{Lemma 5}, the following result is derived.
\begin{theorem}
	\label{Theorem 3}
	\emph{The optimal RB allocation of problem (\ref{SP2-2}) is 	
	\begin{align}	
		\label{So-SP2-2}
\boldsymbol a^*=\underset{\boldsymbol{a}}{\operatorname{argmin}} \sum\nolimits_{k,n}a_{k,n}(\delta_k-v_{k,n}),
	\end{align}
where 
	\begin{align}	
	\label{Co-SP2-2}
	\!\!\! v_{k,n}\!=\!\left\{\begin{array}{lc}
		\!\!-\rho_1\chi^*\frac{\left(I_{n}+W^{U} N_0\right)}{h_k}(2^{\frac{\xi_d+\xi_M}{W^{U} \chi^*}-1}), &\!\!\! \text{if (\ref{TransDelay_hold}) holds} \\
		\delta_k-1, & \!\!\!  \text { otherwise.}   
	\end{array}\right.
\end{align}
}
\end{theorem}

Theorem \ref{Theorem 3} indicates that given $\chi^*$, the optimal RB allocation decision can be determined by solving (\ref{So-SP2-2}). Notably, problem (\ref{So-SP2-2}) is a bipartite graph matching problem with $\mathcal{K}$ and $\mathcal{R}$ being the source set and target set, respectively.
The objective is to find a matching in the graph that minimizes the total sum of weights, i.e., ${\delta_k-v_{k,n}}$ when $\delta_k < v_{k,n}$, and otherwise, ${(\delta_k-v_{k,n})}\rightarrow\infty$.
In view of this, \emph{Kuhn-Munkres} algorithm is utilized to solve problem (\ref{So-SP2-2}) given $\chi^*$\cite{K-M-algorithm}.

Then we assume SU $l$ as the communication straggler.
By fixing $\boldsymbol{a}^*$, problem (\ref{SP2}) is characterized as
\begin{subequations}
	\label{SP2-3}
	\begin{eqnarray}
		\nonumber
		\ \ \  \!\!\!\!\!\!&\!\!\!\!\!\!\underset{\{{\boldsymbol{p}_k}\}_{k\in\mathcal{K}^*}}{\min} \!\!\!\!\!\!&\rho_1 \sum\limits_{n \in \mathcal{R}}\frac{(\xi_d+\xi_M)p_k}{W^{U} \log _2(1+\frac{h_k p_k}{I_n+W^{U} N_0})}  \\ 
		\ \ &&+ \rho_2 \sum\limits_{n \in \mathcal{R}}\frac{\xi_d+\xi_M}{W^{U} \log _2(1+\frac{h_k p_k}{I_n+W^{U} N_0})}  \\
		\label{ap_k_sp2-3}
		&\operatorname{s.t.}& \frac{\left(I_{n_k^*}+W^{U} N_0\right) h_l}{\left(I_{n_l^*}+W^{U} N_0\right) h_k} p_l \leq p_k,  \\
		 &&(\ref{p_k}).
	\end{eqnarray}
\end{subequations}

\begin{lemma}
	\label{Lemma 6}
	\emph{The solution to problem (\ref{SP2-3})	is as follows
	\begin{align}	
		\label{So-SP2-3}
		p_k^*=\left\{\begin{array}{lc}
			{\min}\left\lbrace \underset{k\in\mathcal{K}^*}{\min}\frac{h_k p_k^{max}}{I_{n_k^*}+W^{U} N_0},\hat{p}_l^0\right\rbrace , &\!\!\!\!\! \text { if }k=l \\
			\frac{({I_{n_k^*}+W^{U} N_0})h_l}{({I_{n_l^*}+W^{U} N_0})h_k}p_l^*, &\!\!\!\!\! \text { otherwise }
		\end{array}\right.
	\end{align}
where $\hat{p}_l^0\in(0,c_2]$ is the unique zero point of the function $\hat{Q}(p)=\rho_1c_1((p+1)\log_2({p+1}) \ln2-p)-\rho_2$, which is monotonically increasing for $p\geq0$, and $c_1$ and $c_2$ are
\begin{align}
	\!\!\!\!\!\! c_1\!=\!\sum\limits_{k \in \mathcal{K}^*}\frac{I_{n_k^*}\!+\!W^{U} N_0}{h_k}, 
	~ c_2\!=\!2^{\left(1\!+\!\sqrt{\max \left\{\frac{\rho_2}{\rho_1c_1}, 1\right\}-1}\right) / \ln 2}.\!\!\!\!\!
\end{align}
}
\end{lemma}
\begin{IEEEproof}
	See Appendix E of the technical report \cite{Appendix}.
\end{IEEEproof}

From Lemma \ref{Lemma 6}, the optimal solution of transmission power is derived based on (\ref{So-SP2-3}), given the optimal RB allocation and communication straggler. This differs from Lemma \ref{Lemma 5}, which requires the extra corresponding transmission time $\chi^*$. Therefore, with the RB allocation, the optimal transmission power can be determined by the following theorem.

\begin{theorem}
	\label{Theorem 4}
	\emph{The optimal transmission power of $\mathcal{SP}2$ is 
		\begin{align}
			\label{So-SP2}
			\boldsymbol{p}^*=\underset{p \in \{{p}_l^*\}_{l\in\mathcal{K}}}{\operatorname{argmin}}Q_2(\boldsymbol{a}^*,\boldsymbol{p}),
		\end{align}
		where ${p}_l^*$ indicates the optimal solution of problem (\ref{SP2-3}) given the communication straggler SU $l$.
	}
\end{theorem}

With Theorems \ref{Theorem 3} and \ref{Theorem 4}, $\mathcal{SP}2$ can be solved in an iterative manner.
Concretely, we first determine the RB allocation $a^m$ by solving (\ref{So-SP2-2}) using the \emph{Kuhn-Munkres} algorithm.
Then, the transmission power $p^m$ is computed via $a^m$
 and (\ref{So-SP2}).
 Next, the transmission latency is updated based on $\chi^{m+1}=\max _{k \in \mathcal{K}} \sum_{n \in \mathcal{R}} a_{k, n} T_k^{c o}(a^m,p^m)$ before proceeding to the next iteration.
By incorporating the solutions of $\mathcal{SP}1$ and $\mathcal{SP}2$, 
an efficient joint SU selection and resource allocation algorithm is developed to solve  $\mathcal{P}$, as shown in Algorithm \ref{Optimization algorithm}.

\begin{figure*}[t!]
	\centering
	\begin{subfigure}{0.32\linewidth}
		\centering
		\includegraphics[width=1\linewidth]{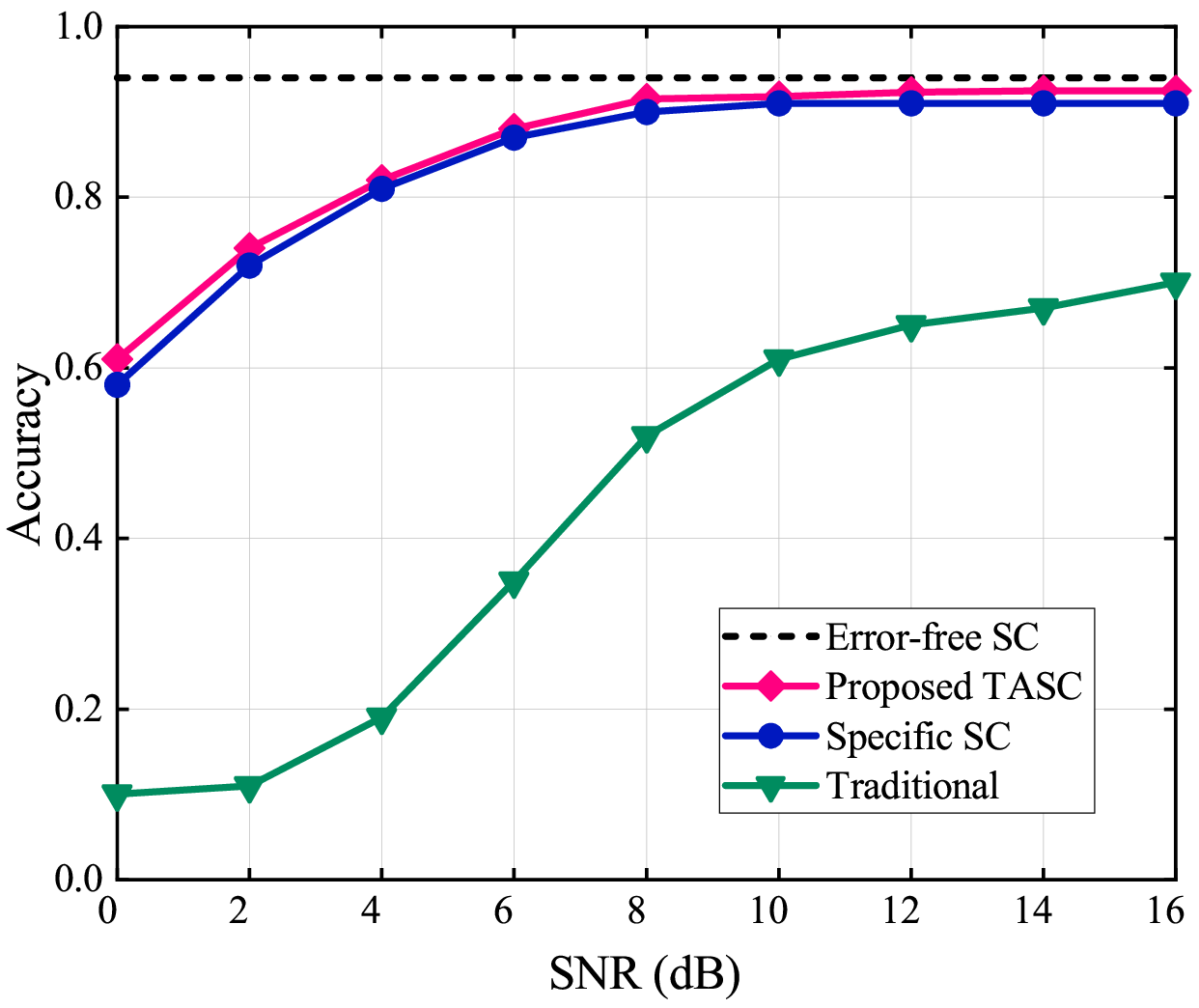}
		\caption{Image classification}
		\label{Rayleigh-classification}
	\end{subfigure}
	\begin{subfigure}{0.32\linewidth}
		\centering
		\includegraphics[width=1\linewidth]{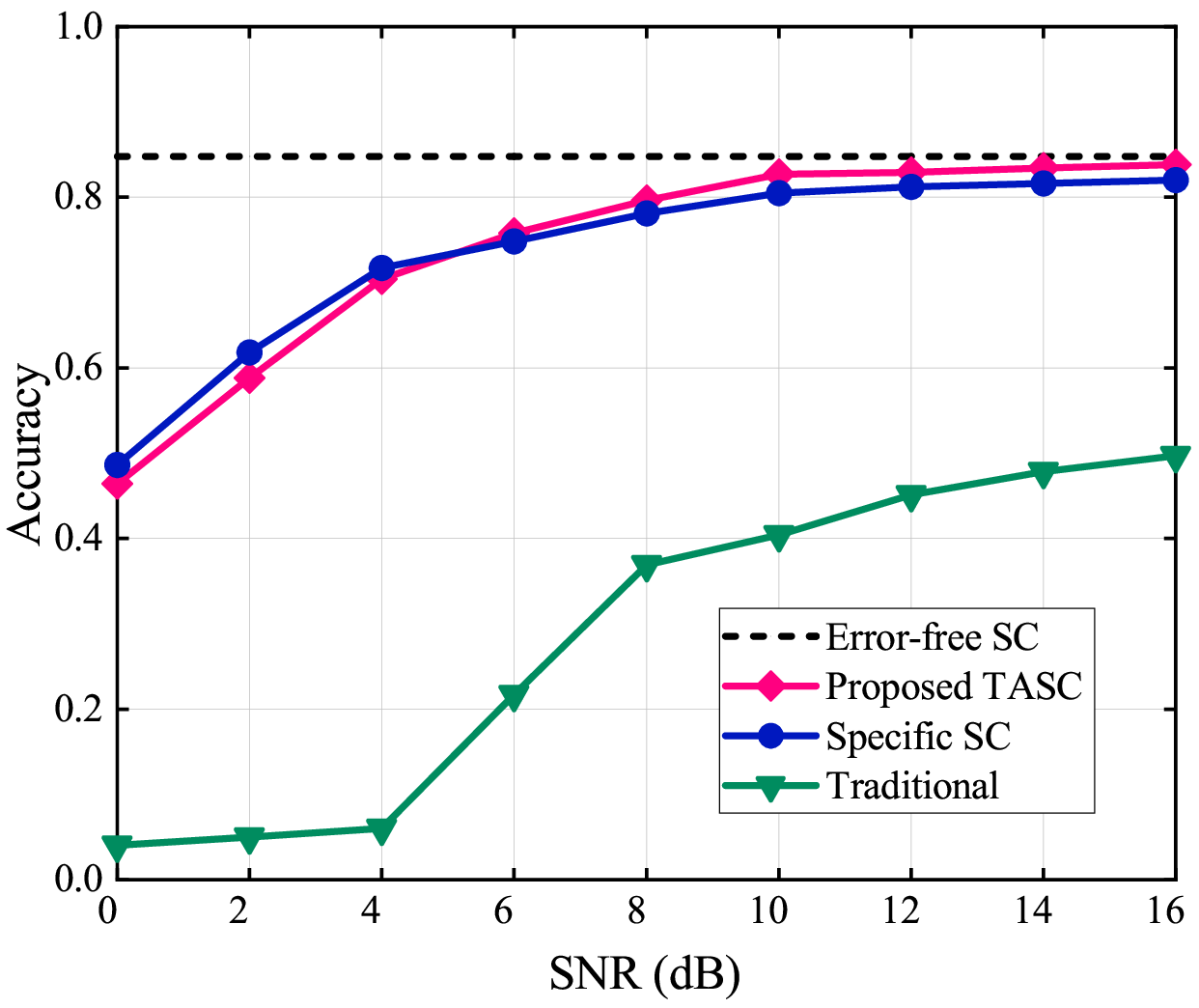}
		\caption{Visual question answering}
		\label{Rayleigh-VQA}
	\end{subfigure}
	\begin{subfigure}{0.32\linewidth}
		\centering
		\includegraphics[width=1\linewidth]{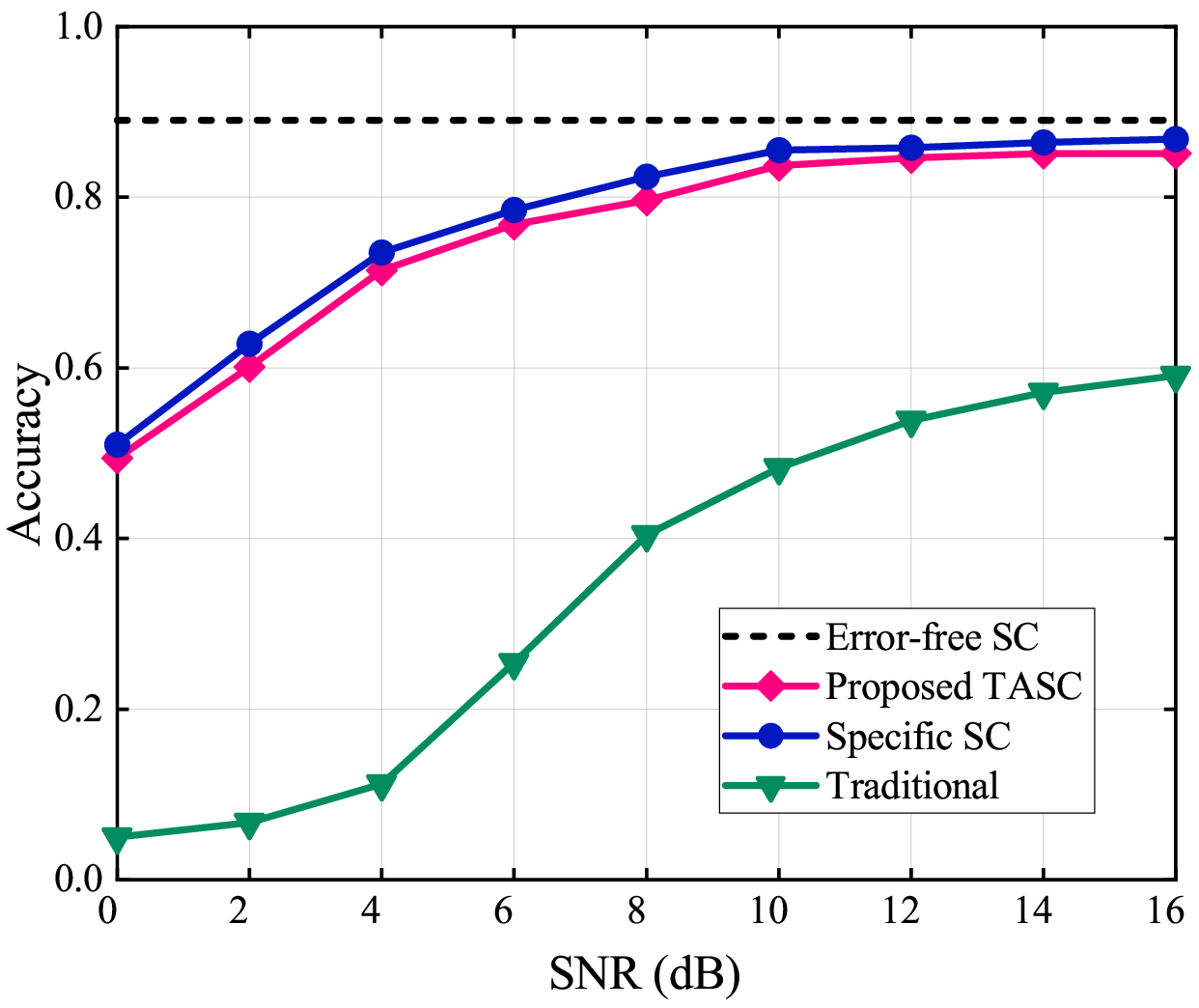}
		\caption{Multimodal sentiment analysis}
		\label{Rayleigh-Multimodal-emotion}
	\end{subfigure}
	\caption{Accuracy performance versus SNR for various tasks under Rayleigh channels.}
	\label{Rayleigh-tasks}
\end{figure*}

\begin{algorithm}[t]
	\caption{The Joint SU Selection and Resource Allocation Algorithm for Solving Problem $\mathcal{P}$}
	\label{Optimization algorithm}
	\textbf{Input:} $ \rho_1, \rho_2, \xi_d, \xi_M, \{h_k\}, \{I_n\};\ \ \ \ \ \ \ \ \ \ \ \ \  $
	\vspace{-4mm}
	\begin{algorithmic}[1]
		\STATE Calculate $\boldsymbol{f}^*$ via (\ref{So-SP1});
		\STATE Initialize iteration $m=0$ and $\chi^0$ via (\ref{TransDelay});
		\WHILE{not  done}
		\STATE Calculate RB decision $a^m$ with $\chi^m$ via (\ref{So-SP2-2});
		\STATE Calculate transmission power $p^m$ with $a^m$ via (\ref{So-SP2});
		\STATE Update $\chi^{m+1}$ with $a^m$ and $p^m$ based on (\ref{TransDelay});		
		\ENDWHILE
	\end{algorithmic}
	\textbf{Output:} $\boldsymbol{a}^*,\boldsymbol{p}^*,\boldsymbol{f}^*$.
\end{algorithm}

\section{Simulation Results}

\subsection{Experimental Setup}
To evaluate the effectiveness of TASC, we conduct experiments on three widely-used tasks.
Specifically, CIFAR-10 dataset is used for image classification task.
The popular CLEVR dataset is selected for the VQA task, and 
the CMU-MOSI dataset is utilized for multimodal sentiment analysis.
The semantic transmitters for different modalities are initialized by the corresponding pretrained Transformer model, such as Vision Transformer, BERT, and Conformer.
The Adam optimized is used with learning rate $2\times10^{-5}$, batch size $64$, and epoch $100$. The Rayleigh fading channels are considered.
In the FML setting, the local dataset of each SU is partitioned into a support set and a query set. Half users are selected for training while the remaining users are used for testing.
The following baselines are considered for comparison.
\begin{figure}[t]
	\centering
	\includegraphics[width=2.6in]{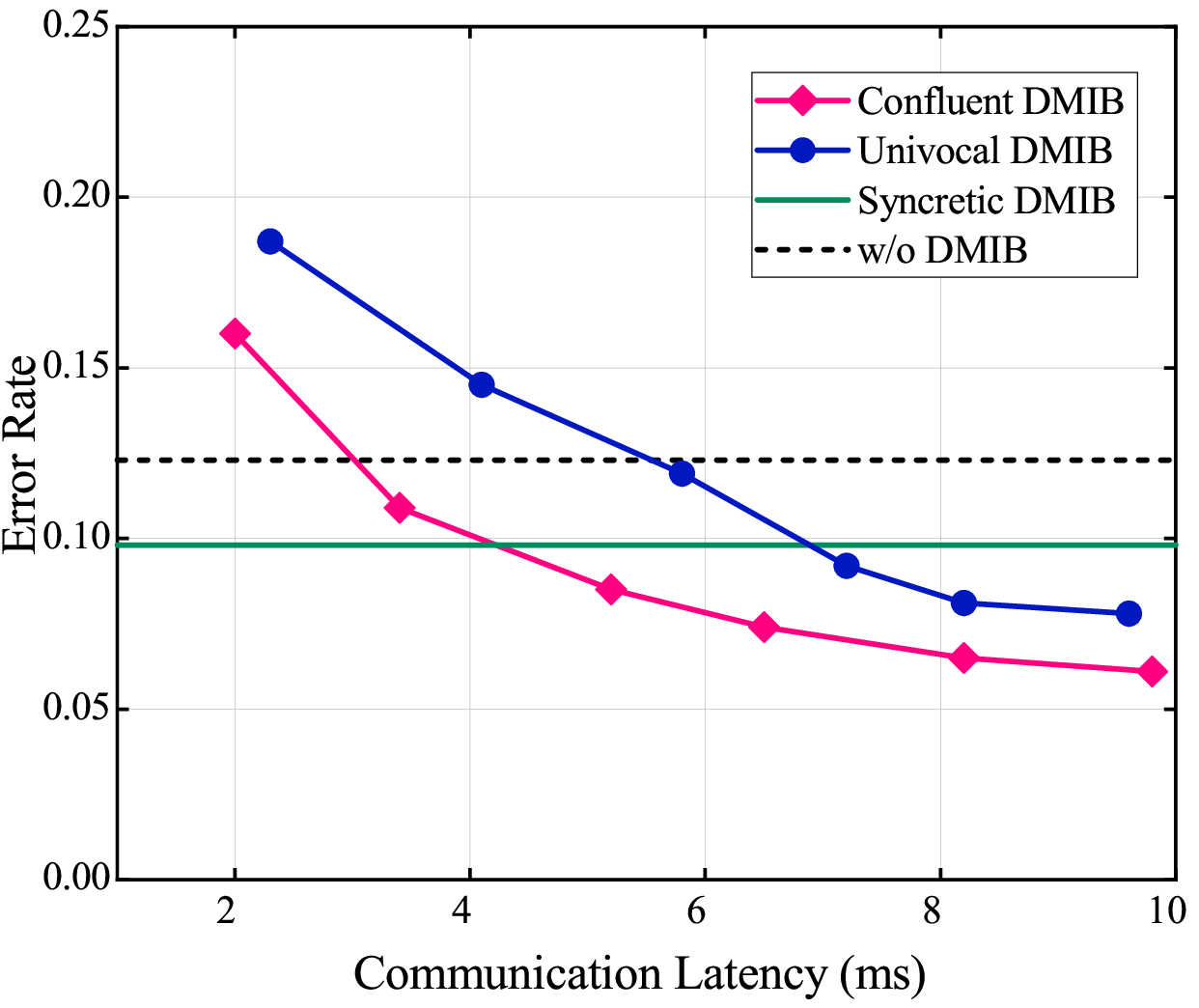}
	\caption{Rate distortion curve for classification task with $8$ dB.}
	\label{rate-distortion}
\end{figure}
\begin{itemize}
\item Error-free SC: In error-free SemCom system (Error-free SC), the inference results are obtained by transmitting noise-free semantic features to the receiver based on separately trained TASC for a specific task.

\item Specific SC: Specific SemCom system (Specific SC) is designed with the same structure as TASC while it is trained for specific tasks separately.

\item Traditional: In traditional source and channel coding, UTF-8, JPEG, AMR-WB, and H.264 are adopted for source coding of text, image, audio, and video, respectively.
LDPC is used for channel coding of image, audio, and video, and the Turbo coding for text channel coding.

\end{itemize}

\begin{figure*}[t!]
	\centering
	\begin{subfigure}{0.32\linewidth}
		\centering
		\includegraphics[width=1\linewidth]{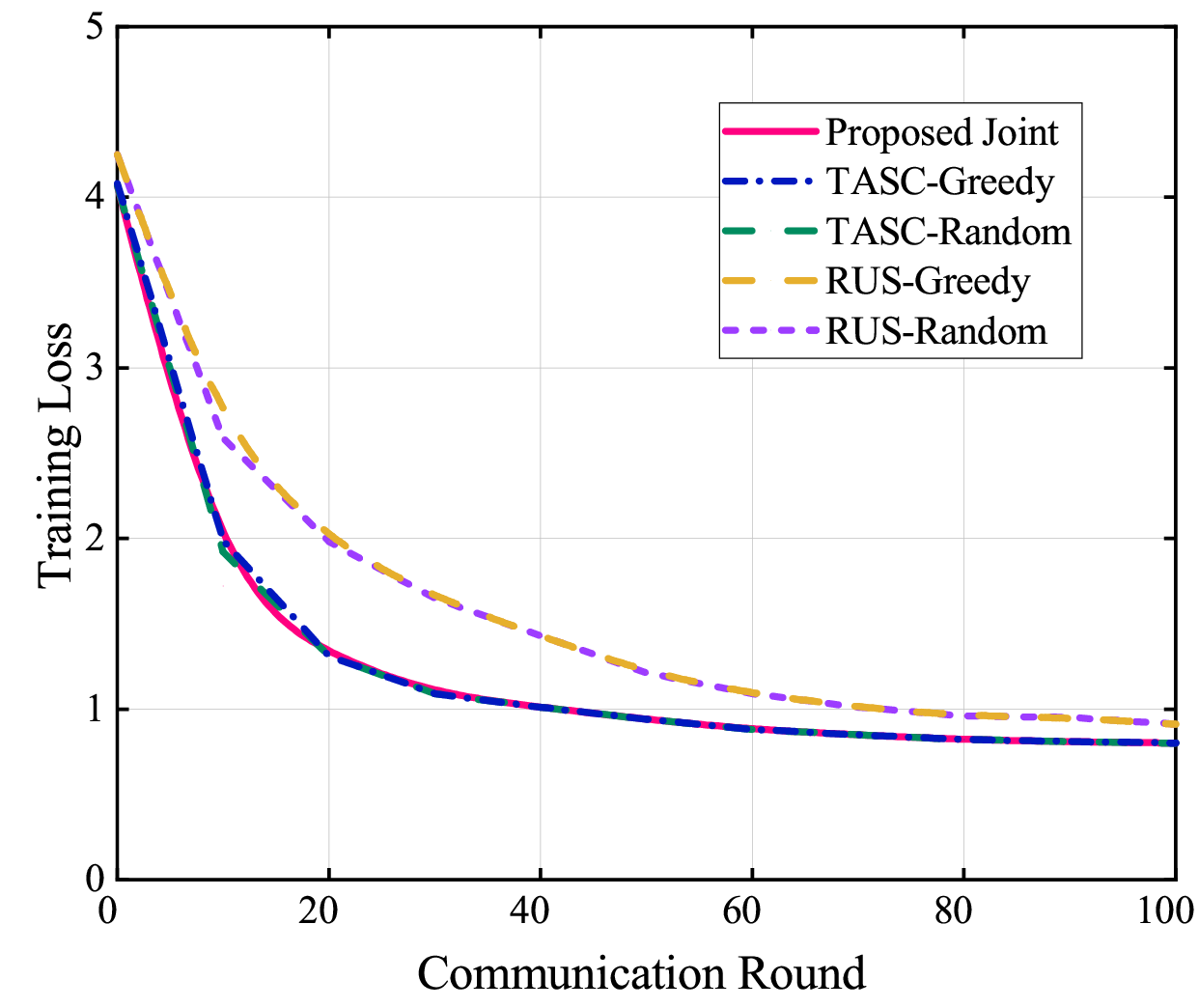}
		\caption{Training loss vs communication round}
		\label{CIFAR-Loss-Rounds}
	\end{subfigure}
	\begin{subfigure}{0.32\linewidth}
		\centering
		\includegraphics[width=1\linewidth]{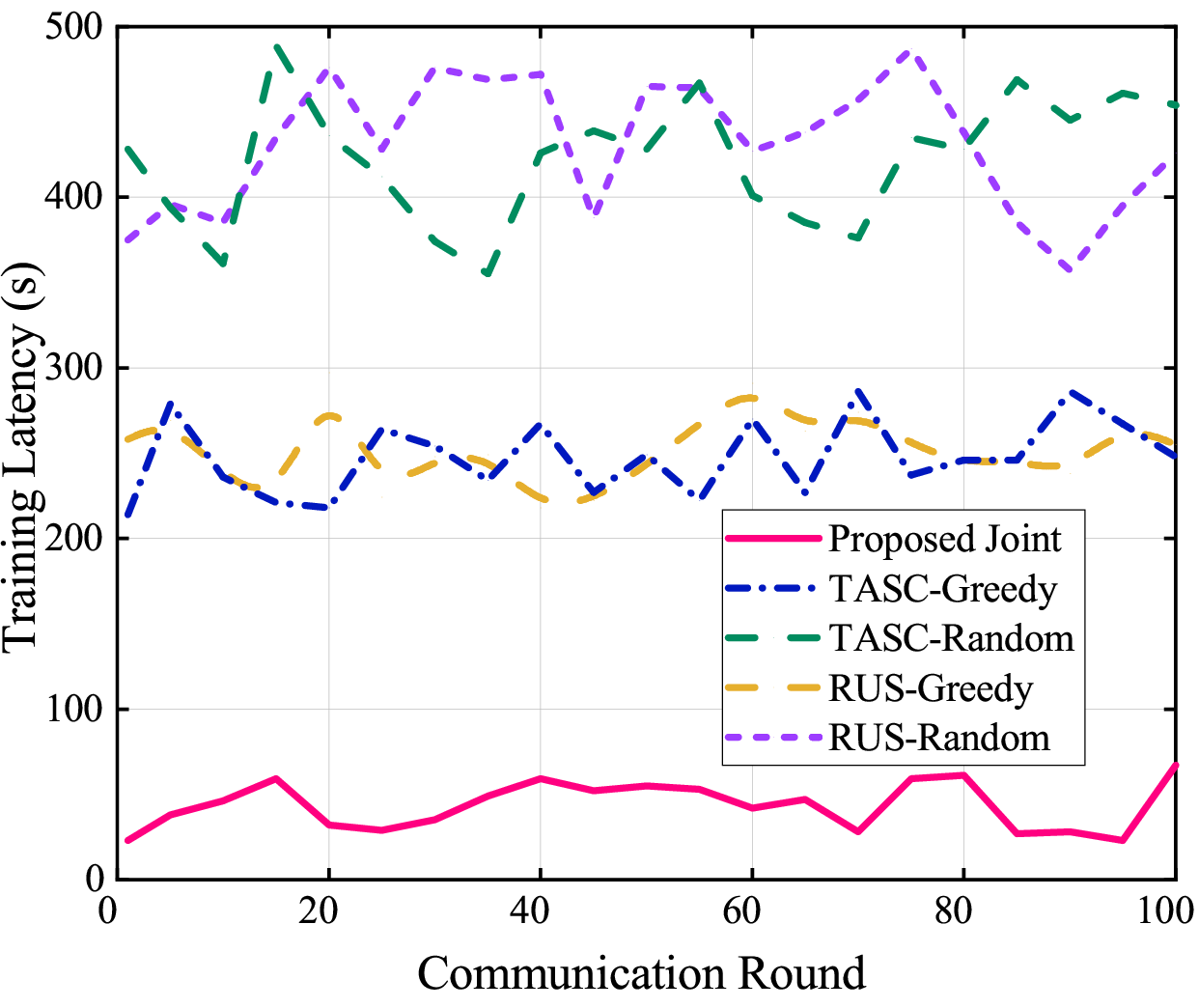}
		\caption{Training time vs communication round}
		\label{CIFAR-Time-Rounds}
	\end{subfigure}
	\begin{subfigure}{0.32\linewidth}
		\centering
		\includegraphics[width=1\linewidth]{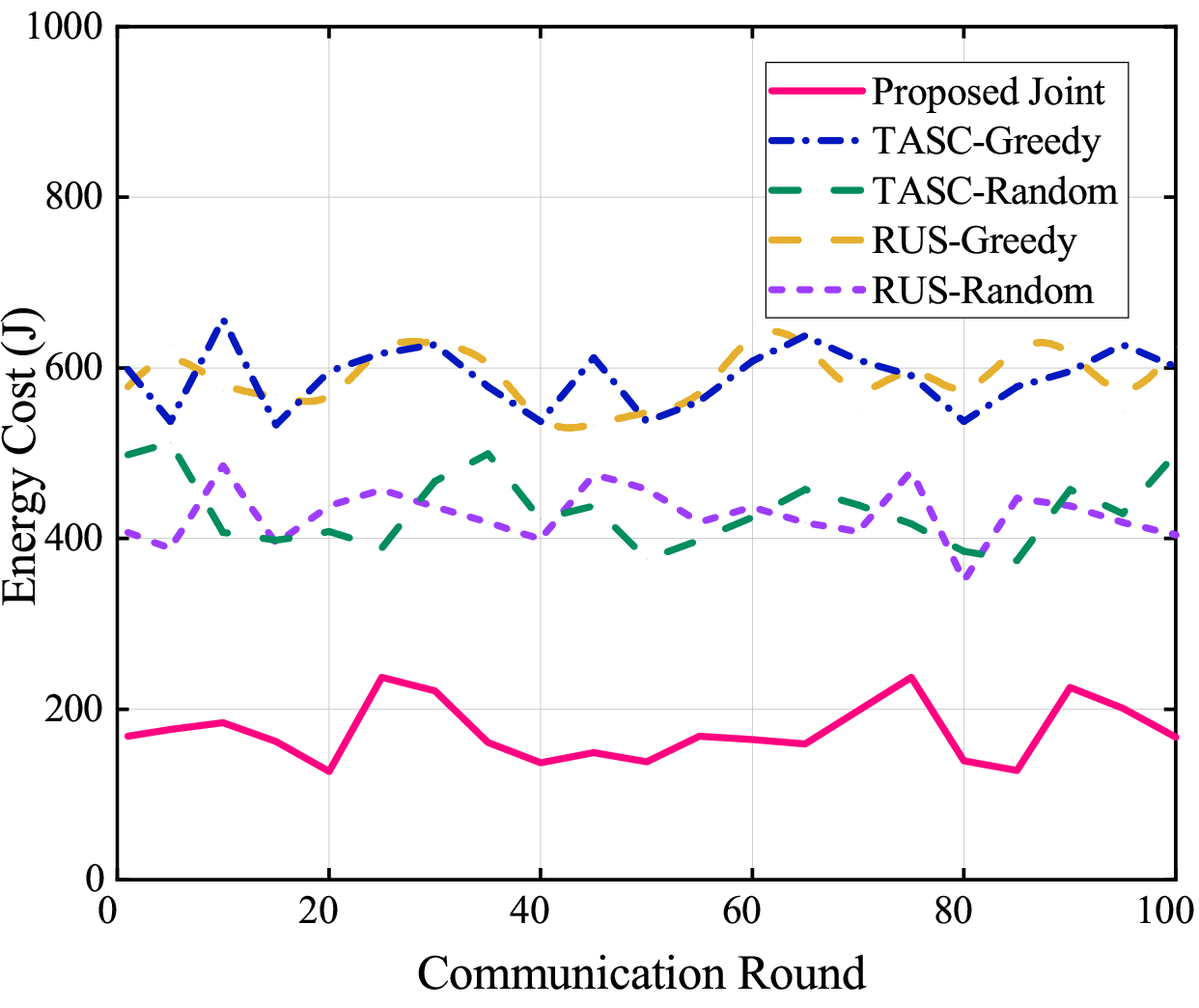}
		\caption{Energy cost vs communication round}
		\label{CIFAR-Energy-Rounds}
	\end{subfigure}
	\caption{Comparison of convergence, training time, and energy cost under image classification task.}
	\label{CIFAR-Opt}
\end{figure*}

\vspace{-4mm}
\subsection{Task Performance}
Fig. \ref{Rayleigh-tasks} depicts the performance of the investigated methods for multiple tasks under various signal-to-noise ratio (SNR) regimes. Note that TASC and Specific SC are trained at the same SNR, i.e., $8$ dB.
It is observed that both the DL-based TASC and Specific SC outperform traditional separate source-channel coding methods, which struggle to effectively mitigate channel distortion, particularly in low SNR conditions.
Besides, the proposed TASC is almost to reach the upper bound at high SNR regimes.
Moreover, TASC demonstrates performance comparable to that of Specific SC across all evaluated tasks, indicating that TASC can efficiently perform multiple diverse tasks while maintaining performance.

Fig. \ref{rate-distortion} illustrates the rate-distortion cures for the image classification task at $8$ dB.
It is seen that the integrated confluent DMIB demonstrates superior performance compared to solely applying either univocal DMIB or syncretic DMIB.
Specifically, for a given latency constraint, it maintains higher classification accuracy, and conversely, for a required accuracy level, it achieves lower latency. This advantage stems from the ability of the proposed DMIB principle to eliminate redundant dimensions in the encoded unimodal features while simultaneously filtering out semantic noise of multimodal representations and preserving task-relevant information, thereby enhancing task-oriented communication efficiency.
 
 \begin{figure}[t]
 	\centering
 	\includegraphics[width=2.6in]{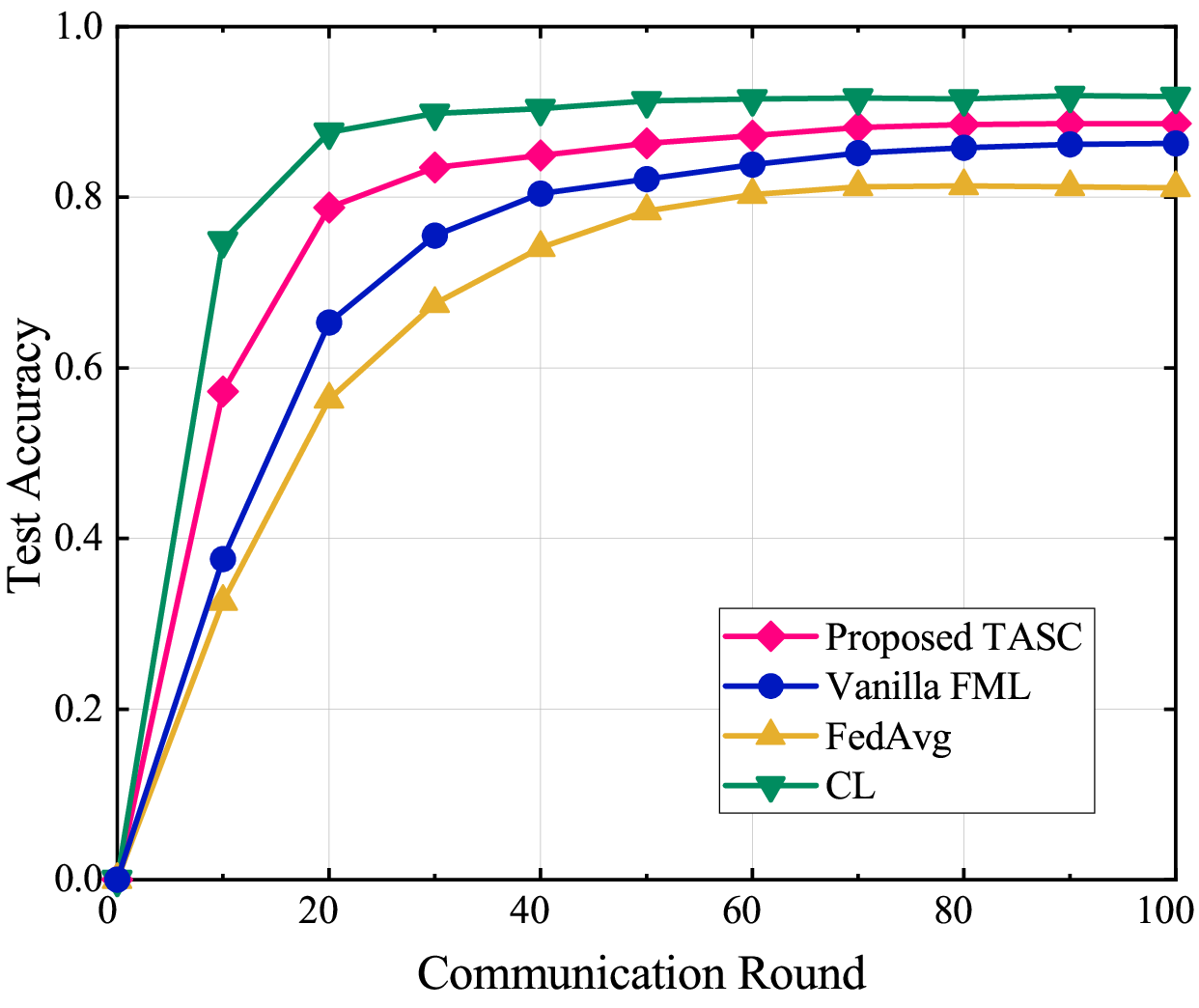}
 	\caption{Test accuracy versus  number of communication rounds.}
 	\label{accuracy-rounds}
 \end{figure}


\vspace{-3mm}
\subsection{Convergence Performance}
Fig. \ref{accuracy-rounds} shows the convergence performance of TASC by evaluating the image classification task.
Denote $\tau_0=1$ and $C=100$.
As shown in Fig. \ref{accuracy-rounds}, it is observed that both the proposed TASC and the vanilla FML achieve better test accuracy than FedAvg with relatively fewer communication rounds.
This is due to the fact that FML can learn a more adaptable initialization that enables faster convergence on tasks.
Moreover, TASC significantly improves the convergence speed with high accuracy in  contrast to the vanilla FML method.
This result clearly demonstrates the effectiveness of our proposed TASC, which minimizes the upper bound of one-round SU loss divergence.
Additionally, TASC employs a shared receiver network at the BS, which eliminates the need for extra parameter aggregation that may introduce aggregating errors, thereby facilitating stable model convergence.

\begin{figure}[t]
	\centering
	\includegraphics[width=2.6in]{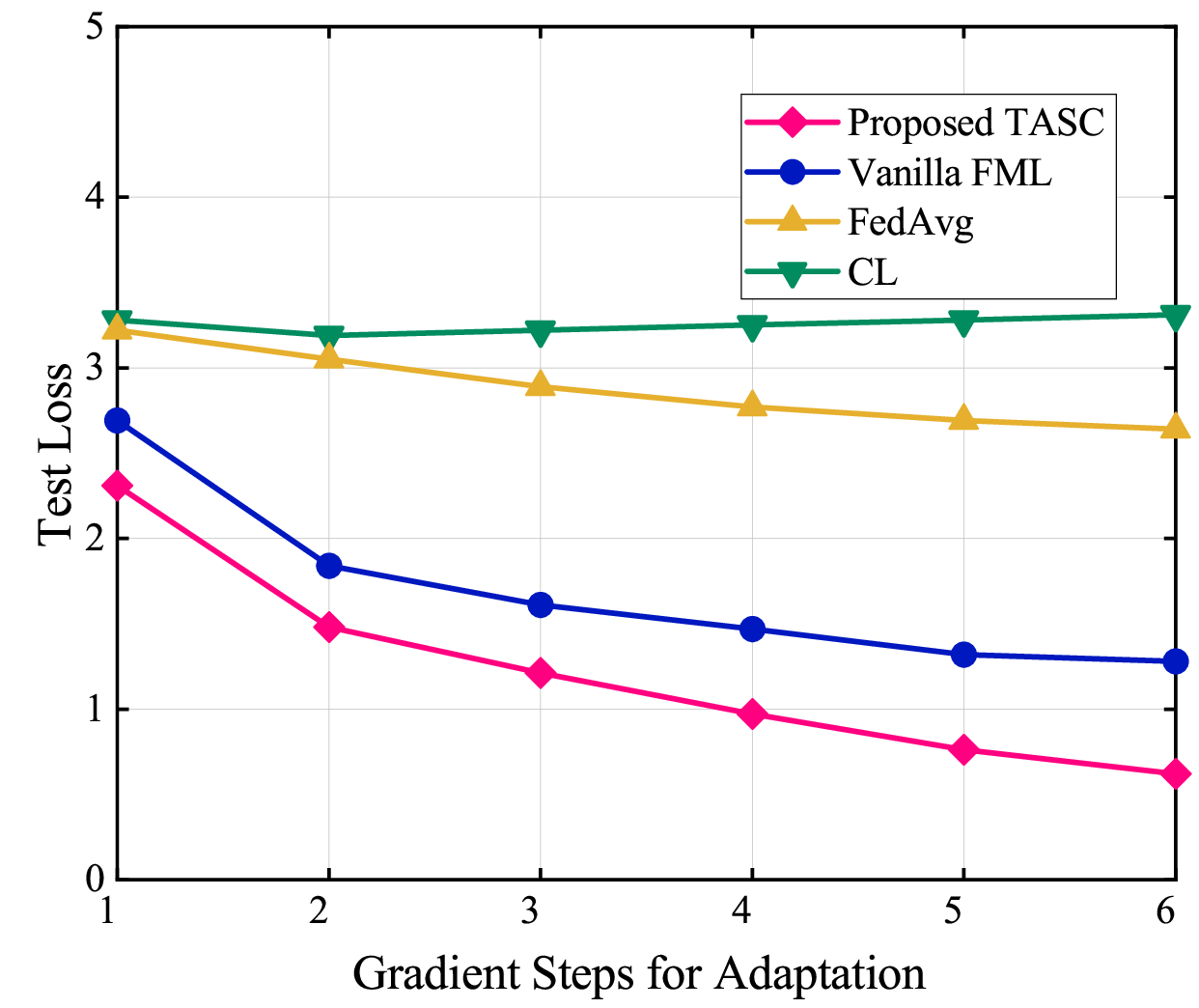}
	\caption{Adaptation performance on image classification task.}
	\label{CIFAR-Adaptation}
\end{figure}


\vspace{-2mm}
\subsection{Effect of Joint Optimization Algorithm}
For the wireless system settings, we set $K=100$, $R=20$, $N_0=-174$ dBm/Hz, $W^U=1$ MHz, $W^D=20$ MHz, $\kappa_B=1/16$ cycles/FLOPs, $f_B=10$ GHz, $p_B=1$ W.
The parameter $f_k$ follows uniform distribution $[0,2]$ GHz, and let $p_k=[0,0.2]$ W, $\varsigma_k=[0,5]\times10^{-28}$, $\kappa_k=[0,1/32]$ cycles/FLOPs.
The inner learning rate $\alpha$ and the meta-learning rate $\beta$ are set to 0.001.
The following baselines are considered: TASC-Greedy, TASC-Random, RUS-Greedy, RUS-Random. TASC is the proposed SU selection strategy by solving (\ref{P-DQ}) and RUS is to select SUs uniformly at random. Greedy and Random are strategies to decide CPU-cycle frequencies, RB allocation, and transmission
power.
From Fig. \ref{CIFAR-Opt}, it shows that the proposed joint optimization algorithm significantly reduces both energy cost and training latency in contrast to baselines.
It is also observed that the Greedy algorithm does not always outperform the Random.
This is because that the weight parameters $\rho_1$ and $\rho_2$ influence the energy cost and training latency.
When $\rho_1=\rho_2$, the Greedy algorithm prioritizes minimizing training latency over energy cost.
Besides, both Greedy and Random are lack of effective joint optimization, leading to performance degradation.
However, the proposed algorithm outperforms baselines by minimizing the objective $\mathcal{P}$, which mitigates the aforementioned limitations, resulting in improved performance in resource-constrained wireless systems.


\vspace{-2mm}
\subsection{Performance of Rapid Adaptation}

To demonstrate the adaptation performance of TASC for brand-new users/data,
we present results by varying the number of gradient steps.
As illustrated in Fig. \ref{CIFAR-Adaptation}, it can be observed that TASC obtains significantly better adaptation performance by achieving about 51\% and 59\% decreases in terms of test loss respectively compared to the vanilla FML and FedAvg.
This performance gap becomes more pronounced as the number of gradient steps increases.
Moreover, it is indicated that FedAvg and CL suffer from over-fitting issues, whereas the FML methods continue to improve with the increasing gradient steps without over-fitting.
Therefore, the proposed TASC can adapt to new environments through one or a few gradient descent steps without requiring retraining.

\section{Conclusion}
In this paper, we proposed a task-agnostic SemCom framework named TASC over wireless networks.
Guided by the IB theory, we developed a novel DMIB principle to capture minimal sufficient unimodal and multimodal semantic representations by removing redundancy and semantic noise while retaining task-related information.
Moreover, we proposed an adaptive semantic transmission approach to adjust the transmission dimensions under dynamic channel conditions.
Then, TASC is trained based on FML for quick adaptation and generalization.
We also conducted convergence analysis, implying the importance of SU selection to convergence.
Based on the results, we formulated a resource management problem to optimize the tradeoff among convergence, training latency, and energy cost, and developed a joint optimization algorithm to solve it.
Extensive simulation results verified that the
proposed TASC outperforms the task-specific SemCom and achieves a better rate-distortion tradeoff.
We also evaluated the superiority of TASC over baselines in
terms of convergence speed, training latency, and energy cost.
Benefited from FML, TASC achieves rapid adaptation over only a few SGD steps.


\onecolumn
\appendices

\section{Proof of Theorem \ref{theorem 1}} 
\label{proof_of_theorem_1}
For brevity, let ${\boldsymbol{\theta}}_{t, \tau}^k={\boldsymbol{\theta}}_k({t, \tau})$.
We denote that ${\boldsymbol{\theta}}_{t, \tau}=1 / K \sum_{k=1}^K \boldsymbol{\theta}_{t, \tau}^k$, and $\bar{\boldsymbol{\theta}}_{t, \tau}=1 / K^t \sum_{k=1}^{K^t} \boldsymbol{\theta}_{t, \tau}^k$.
Based on Lemma \ref{Lemma 1}, $L(\cdot)$ is smooth with Lipschitz parameter $H_L$, and thus, we have
\begin{align}
	\nonumber
	& L\left(\bar{\boldsymbol{\theta}}_{t+1, \tau+1}\right) \\	\nonumber
	& \leq L\left(\bar{\boldsymbol{\theta}}_{t+1, \tau}\right)+\nabla L\left(\bar{\boldsymbol{\theta}}_{t+1, \tau}\right)^{\top}\left(\bar{\boldsymbol{\theta}}_{t+1, \tau+1}-\bar{\boldsymbol{\theta}}_{t+1, \tau}\right)+\frac{H_L}{2}\left\|\bar{\boldsymbol{\theta}}_{t+1, \tau+1}-\bar{\boldsymbol{\theta}}_{t+1, \tau}\right\|^2 \\	
	& \leq L\left(\bar{\boldsymbol{\theta}}_{t+1, \tau}\right)-\beta \nabla L\left(\bar{\boldsymbol{\theta}}_{t+1, \tau}\right)^{\top}\left(\frac{1}{K^t} \sum_{k \in \mathcal{K}_t} \tilde{\nabla} L_k\left(\bar{\boldsymbol{\theta}}_{t+1, \tau}^k\right)\right)+\frac{H_L}{2} \beta^2\left\|\frac{1}{K^t} \sum_{K \in \mathcal{K}_t} \tilde{\nabla} L_k\left(\bar{\boldsymbol{\theta}}_{t+1, \tau}^k\right)\right\|^2,
\end{align}
where $\mathcal K^t$ is the set of selected SUs with size $K^t$ in the $t$-th round.
By taking expectation, we have
\begin{align}
		\nonumber
	 \mathbb{E}\left[L\left(\bar{\boldsymbol{\theta}}_{t+1, \tau+1}\right)\right] \leq &\mathbb{E}\left[L\left(\bar{\boldsymbol{\theta}}_{t+1, \tau}\right)\right]-\beta \mathbb{E}\left[\nabla L\left(\bar{\boldsymbol{\theta}}_{t+1, \tau}\right)^{\top}\left(\frac{1}{K^t} \sum_{k \in \mathcal{K}^t} \tilde{\nabla} L_k\left({\boldsymbol{\theta}}_{t+1, \tau}^k\right)\right)\right] \\ \label{proof_thoorem1_L}
	&+ \frac{H_L}{2} \beta^2 \mathbb{E}\left[\left\|\frac{1}{K^t} \sum_{k \in \mathcal{K}^t} \tilde{\nabla} L_k\left({\boldsymbol{\theta}}_{t+1, \tau}^k\right)\right\|^2\right].
\end{align}
Next, note that
\begin{align}
\frac{1}{K^t} \sum_{k \in \mathcal{K}^t} \tilde{\nabla} L_k\left({\boldsymbol{\theta}}_{t+1, \tau}^k\right)=C_1+C_2+C_3+\frac{1}{K^t} \sum_{k \in \mathcal{K}^t} \nabla L_k\left(\bar{\boldsymbol{\theta}}_{t+1, \tau}\right),
\end{align}
where
\begin{align}
&	C_1=\frac{1}{K^t} \sum_{k \in \mathcal{K}^t} \left(\tilde{\nabla} L_k\left(\boldsymbol{\theta}_{t+1, \tau}^k\right)-\nabla L_k\left(\boldsymbol{\theta}_{t+1, \tau}^k\right)\right), \\
&	C_2=\frac{1}{K^t} \sum_{k \in \mathcal{K}^t} \left({\nabla} L_k\left(\boldsymbol{\theta}_{t+1, \tau}^k\right)-\nabla L_k\left(\boldsymbol{\theta}_{t+1, \tau}\right)\right),     \\
&	C_3=\frac{1}{K^t} \sum_{k \in \mathcal{K}^t} \left({\nabla} L_k\left(\boldsymbol{\theta}_{t+1, \tau}\right)-\nabla L_k\left(\bar{\boldsymbol{\theta}}_{t+1, \tau}\right)\right) .
\end{align}
By bounding the moments of $C_1, C_2$, and $C_3$, we can obtain
\begin{align}
	\mathbb{E}\left[\|C_1\|^2\right]\leq \sigma_L^2.
\end{align}
\begin{align}
		\nonumber
	\mathbb{E}\left[\|C_2\|^2\right] &  \leq H_L^2 \mathbb{E}\left[\frac{1}{K} \sum_{k=1}^K\left\|\boldsymbol{\theta}_{t, \tau}^k-\boldsymbol{\theta}_{t, \tau}^k\right\|^2\right] \\
	& \leq 35 \beta^2 H_L^2 \tau_0(\tau_0-1)\left(2 \sigma_L^2+\mu_L^2\right).
\end{align}
\begin{align}
		\nonumber
		\mathbb{E}\|C_3\|^2 & \leq \frac{1}{K^t}	\mathbb{E}\left[ \sum_{k \in \mathcal{K}^t}\left\|\nabla L_k\left({\boldsymbol{ \theta}}_{t+1, \tau}\right)-\nabla L_k\left(\bar{\boldsymbol{ \theta}}_{t+1, \tau}\right)\right\|^2\right] \\
	& \leq  \frac{35(1-\frac{K^t}{K}) \beta^2 H_L^2 \tau(\tau-1)\left(2 \sigma_L^2+\mu_L^2\right)}{K^t-\frac{K^t}{K}}.
\end{align}
Next, we manage to lower bound the second term in (\ref{proof_thoorem1_L}), and we have
\begin{align}
		\nonumber
	\mathbb{E} & {\left[\nabla L\left(\bar{\boldsymbol{\theta }}_{t+1, \tau}\right)^{\top}\left(\frac{1}{K^t} \sum_{k \in \mathcal{K}^t} \tilde{\nabla} L_k\left(\boldsymbol{\theta }_{t+1, \tau}^k\right)\right)\right] } \\	\nonumber
	& =\mathbb{E}\left[\nabla L\left(\bar{\boldsymbol{\theta }}_{t+1, \tau}\right)^{\top}\left(C_1+C_2+C_3+\frac{1}{K^t} \sum_{k \in \mathcal{K}^t} \nabla L_k\left(\bar{\boldsymbol{\theta }}_{t+1, \tau}\right)\right)\right] \\	\nonumber
	& \left.\geq \mathbb{E}\left[\nabla L\left(\bar{\boldsymbol{\theta }}_{t+1, \tau}\right)^{\top}\left(\frac{1}{K^t} \sum_{k \in \mathcal{K}^t} \nabla L_k\left(\bar{\boldsymbol{\theta }}_{t+1, \tau}\right)\right)\right]-\left\|\mathbb{E}\left[\nabla L\left(\bar{\boldsymbol{\theta }}_{t+1, \tau}\right)^{\top} C_1\right]\right] \right\| \\	\nonumber
	& ~~~ -\frac{1}{4} \mathbb{E}\left[\left\|\nabla L\left(\bar{\boldsymbol{\theta }}_{t+1, \tau}\right)\right\|^2\right]-\mathbb{E}\left[\|C_2+C_3\|^2\right] \\ \label{proof_thoorem1_L_sec}
	& \quad \geq \frac{1}{2} \mathbb{E}\left[\left\|\nabla L\left(\bar{\boldsymbol{\theta }}_{t+1,\tau}\right)\right\|^2\right]-140 \beta^2 H_L^2 \tau_0(\tau_0-1)\left(2 \sigma_L^2+\mu_L^2\right)-\frac{4 \alpha^2 H^2 \sigma_G^2}{D}.
\end{align}
Then, we represent the upper bound for the third term in (\ref{proof_thoorem1_L}):
\begin{align}
	\nonumber
	& \mathbb{E}\left[\left\|\frac{1}{K^t} \sum_{k \in \mathcal{K}^t} \tilde{\nabla} L_k\left(\boldsymbol{\theta}_{t+1, \tau}^k\right)\right\|^2\right] \\	\nonumber
	& \leq 2 \mathbb{E}\left[\left\|\frac{1}{K^t} \sum_{k \in \mathcal{K}^t} \nabla L_k\left(\bar{\boldsymbol{\theta}}_{t+1, \tau}\right)\right\|^2\right]+4 \sigma_L^2+560 \beta^2 H_L^2 \tau_0(\tau_0-1)\left(2 \sigma_L^2+\mu_L^2\right)\\ \label{proof_thoorem1_L_thir}
	&\leq 2 \mathbb{E}\left[\left\|\nabla L\left(\bar{\boldsymbol{ \theta}}_{t+1, \tau}\right)\right\|^2\right]+\frac{2 \mu_L^2(1-\frac{K^t}{K})}{K^t-\frac{K^t}{K}}+4 \sigma_L^2+560 \beta^2 H_L^2 \tau_0(\tau_0-1)\left(2 \sigma_L^2+\mu_L^2\right).
\end{align}
By substituting (\ref{proof_thoorem1_L_sec}) and (\ref{proof_thoorem1_L_thir}) in (\ref{proof_thoorem1_L}), we have
\begin{align}
	\label{proof_thoorem1_total}
 \mathbb{E}\left[L\left(\bar{\boldsymbol{ \theta}}_{t+1, \tau+1}\right)\right]  \leq \mathbb{E}\left[L\left(\bar{\boldsymbol{ \theta}}_{t+1, \tau}\right)\right]-\frac{\beta}{4} \mathbb{E}\left[\left\|\nabla L\left(\bar{\boldsymbol{ \theta}}_{t+1, \tau}\right)\right\|^2\right]+\beta \sigma_T^2,
\end{align}
where
\begin{align}
	\sigma_T^2=280\left(\beta H_L\right)^2 \tau_0(\tau_0-1)\left(2 \sigma_L^2+\mu_L^2\right)+\beta H_L\left(2 \sigma_L^2+\frac{\mu_L^2(1-\frac{K^t}{K})}{K^t-\frac{K^t}{K}}\right)+\frac{4 \alpha^2 H^2 \sigma_G^2}{D}.
\end{align}
Finally, summarizing (\ref{proof_thoorem1_total}) from $\tau=0, \ldots, \tau_0-1$ and $t=0, \ldots, C-1$, we can obtain
\begin{align}
	\mathbb{E}\left[L\left(\boldsymbol{ \theta}_C\right)\right] \leq L\left(\boldsymbol{ \theta}_0\right)-\frac{\beta \tau_0 C}{4}\left(\frac{1}{\tau_0 C} \sum_{t=0}^{C-1} \sum_{\tau=0}^{\tau_0-1} E\left[\left\|\nabla L\left(\bar{\boldsymbol{ \theta}}_{t+1, \tau}\right)\right\|^2\right]\right)+\beta \tau_0 C \sigma_T^2.
\end{align}
Hence, we have
\begin{align}
	\nonumber
	\frac{1}{\tau_0 C} \sum_{t=0}^{C-1} \sum_{t=0}^{\tau_0-1} E\left[\left\|\nabla L\left(\bar{\boldsymbol{ \theta}}_{t+1, \tau}\right)\right\|^2\right] & \leq \frac{4}{\beta \tau_0 C}\left(L\left(\boldsymbol{ \theta}_0\right)-\mathbb{E}\left[L\left(\boldsymbol{ \theta}_C\right)\right]+\beta \tau_0 C \sigma_T^2\right) \\
	& \leq \frac{4\left(L\left(\boldsymbol{ \theta}_0\right)-L\left(\boldsymbol{\theta}_\epsilon\right)\right)}{\beta \tau_0 C}+4 {\sigma}_T^2
\end{align}
If we set $\tau_0=1$, we obtain the desired result.

\section{Proof of Corollary \ref{corollary 1}} 
\label{proof_of_corollary_1}
Based on Lemma 1, for any $\boldsymbol\theta_1, \boldsymbol\theta_2$, we have
\begin{align}
\label{proof_cor_smooth}
	L\left(\boldsymbol\theta_2\right) \leq L\left(\boldsymbol\theta_1\right)+\nabla L\left(\boldsymbol\theta_1\right)^{\top}\left(\boldsymbol\theta_2-\boldsymbol\theta_1\right)+\frac{H_L}{2}\left\|\boldsymbol\theta_2-\boldsymbol\theta_1\right\|^2.
\end{align}
Combining (\ref{proof_cor_smooth}) with (\ref{outer update}), we have
\begin{align}
	\nonumber
	L\left(\boldsymbol\theta_{t,\tau+1}\right) & \leq L\left(\boldsymbol\theta_{t,\tau}\right)+\nabla L\left(\boldsymbol\theta_{t,\tau}\right)^{\top}\left(\boldsymbol\theta_{t,\tau+1}-\boldsymbol\theta_{t,\tau}\right)+\frac{H_L}{2}\left\|\boldsymbol\theta_{t,\tau+1}-\boldsymbol\theta_{t,\tau}\right\|^2 \\
	& =L\left(\boldsymbol\theta_{t,\tau}\right)-\beta \nabla L\left(\boldsymbol\theta_{t,\tau}\right)^{\top}\left(\frac{1}{K^t} \sum_{k \in \mathcal{K}^t} \tilde{\nabla} L_k\left(\boldsymbol\theta_{t,\tau}\right)\right)+\frac{H_L \beta^2}{2}\left\|\frac{1}{K^t} \sum_{k \in \mathcal{K}^t} \tilde{\nabla} L_k\left(\boldsymbol\theta_{t,\tau}\right)\right\|^2.
\end{align}
We denote
\begin{align}
	G_{t,\tau}=\beta \nabla L\left(\boldsymbol\theta_{t,\tau}\right)^{\top}\left(\frac{1}{K^t} \sum_{k \in \mathcal{K}^t} \tilde{\nabla} L_k\left(\boldsymbol\theta_{t,\tau}\right)\right)-\frac{H_L \beta^2}{2}\left\|\frac{1}{K^t} \sum_{k \in \mathcal{K}^t} \tilde{\nabla} L_k\left(\boldsymbol\theta_{t,\tau}\right)\right\|^2.
\end{align}
$\nabla L\left(\boldsymbol\theta_{t,\tau}\right)$ can be rewritten as
\begin{align}
\nabla L\left(\boldsymbol\theta_{t,\tau}\right)=\underbrace{\nabla L\left(\boldsymbol\theta_{t, \tau}\right)-\nabla L\left(\boldsymbol\theta_{t,\tau}^k\right)}_{E_1}+\underbrace{\nabla L\left(\boldsymbol\theta_{t,\tau}^k\right)-\nabla L_k\left(\boldsymbol\theta_{t,\tau}^k\right)}_{E_2}+\underbrace{\nabla L_k\left(\boldsymbol\theta_{t,\tau}^k\right)-\tilde{\nabla} L_k\left(\boldsymbol\theta_{t,\tau}^k\right)}_{E_3}+\tilde{\nabla} L_k\left(\boldsymbol\theta_{t,\tau}^k\right),
\end{align}
where
 $\mathbb{E}\left[\left\|E_1\right\|^2\right]$,  $\mathbb{E}\left[\left\|E_2\right\|^2\right]$, and $\mathbb{E}\left[\left\|E_3\right\|^2\right]$ are bounded by
\begin{align}
	&\mathbb{E}\left[\left\|E_1\right\|^2\right]  \leq  35 \beta^2\tau_0^2(\mu_G^2+2\sigma_F^2),    \\
	& \mathbb{E}\left[\left\|E_2\right\|^2\right] \leq(1+\alpha H)^2 \mu_G+\alpha \Phi \mu_H, \\
	& \mathbb{E}\left[\left\|E_3\right\|^2\right] \leq \sigma_{L}^2.
\end{align}
Then $\mathbb{E}\left[G_{t,\tau}\right]$ can be bounds as
\begin{align}
	\nonumber
\mathbb{E}\left[G_{t,\tau}\right]\geq &\beta \mathbb{E}[\frac{1}{K^t} \sum_{k \in \mathcal{K}^t}\left(1-\frac{H_L \beta}{2}\right)\left\|\tilde{\nabla}L_k\left(\boldsymbol\theta_{t,\tau}\right)\right\|^2\\
&-\left(\sqrt{(1+\alpha H)^2 \mu_G+\alpha \Phi \mu_H}+\sigma_{L}+\beta
\sqrt{35 \beta^2\tau_0^2(\mu_G^2+2\sigma_L^2)}\right) \sqrt{\mathbb{E}\left[\left\|\bar{\nabla} L_k\left(\boldsymbol\theta_{t,\tau}\right)\right\|^2 \mid \mathcal{K}^t\right]}].
\end{align}
Hence we have
\begin{align}
	\nonumber
	 \mathbb{E}\left[L\left(\boldsymbol\theta_{t+1}\right)-L\left(\boldsymbol\theta_{t}\right)\right] &=\mathbb{E}\left[\sum_{\tau=0}^{\tau_0-1} L\left(\boldsymbol\theta_{t, \tau+1}\right)-L\left(\boldsymbol\theta_{t, \tau}\right)\right] \\ \nonumber
	 &\leq -\sum_{\tau=0}^{\tau_0-1} \mathbb{E}\left[G_{t, \tau}\right] \\
	 & \leq \frac{\beta}{2} \mathbb{E}\left[\frac{ 1 } {K^t} \sum _ { k \in \mathcal { K }^{ t } } ((\eta_1+\frac{\eta_2}{\sqrt{D_k}}) \|\tilde{\nabla} L_k\left(\boldsymbol\theta_k(t)\right)\| 
 -\|\tilde{\nabla} L_k\left(\boldsymbol\theta_k(t)\right)\|^2)\right],
\end{align}
where
\begin{align}
	\label{eta1}
	&\eta_1\geq  \sqrt{16\mu_G+4\alpha\Phi\mu_H}+\beta\sqrt{140(\mu_G^2+2\sigma_L^2)}, \\ \label{eta2}
	& \eta_2 \geq 24\sigma_G^2(4+\alpha^2\sigma_H^2)+6\alpha^2\Phi^2\sigma_H^2.
\end{align}

\section{Proof of Lemma \ref{Lemma 4}}
\label{proof_of_lemma_4}
Since $l$ is the straggler among SUs, $\mathcal{SP}$1 can be represented as
 \begin{subequations}
 	\label{proof_4_1}
	\begin{eqnarray}
		\label{SP1-1-function}
		& \underset{\boldsymbol{f}}{\min}   & \rho_1 \sum_{k \in \mathcal{K}}\varsigma_k  \kappa_k \gamma_k D_k f_k^2  +\rho_2 \frac{ \kappa_l \gamma_l D_l}{f_l}  \\
		\label{f_k_sp1}
		&\operatorname{s.t.} &\frac{ \kappa_k \gamma_k D_k f_l}{\kappa_l \gamma_l D_l} \leq {f_k},  \quad \forall k \in \mathcal{K} / l, \\
		\label{f_k_sp2}
		&&  (\ref{f_k}).
	\end{eqnarray}
\end{subequations}
By fixing $f_l$, the optimal CPU frequency $f_k^*$ of the problem (\ref{proof_4_1}) can be obtained by solving the following decomposed convex optimization problem
 \begin{subequations}
  	\label{proof_4_2}	
\begin{eqnarray}
		& \underset{\boldsymbol{f}}{\min}   & \rho_1 \sum_{k \in \mathcal{K}}\varsigma_k  \kappa_k \gamma_k D_k f_k^2    \\
		&\operatorname{s.t.} &\frac{ \kappa_k \gamma_k D_k f_l}{\kappa_l \gamma_l D_l} \leq {f_k},  \\
&&  (\ref{f_k}).
\end{eqnarray}
\end{subequations}
If $f_l \leq \frac{ \kappa_k \gamma_k D_k f_k^{\text{max}}}{\kappa_l \gamma_l D_l}$, the optimal solution of problem (\ref{proof_4_2}) is denoted by
\begin{align}
	  	\label{proof_so_1}
	f_k^*=\frac{ \kappa_k \gamma_k D_k f_l}{\kappa_l \gamma_l D_l} 
\end{align}
Then we substitute $f_k=f_k^*$ in problem (\ref{proof_4_1}) and we have
 \begin{subequations} 
  	\label{proof_4_3}
\begin{eqnarray}	
	\min _{f_l} ~~& g_1\left(f_l\right) = \underbrace{\rho_1\left(\sum_{k \in \mathcal{K} / l} \frac{\varsigma_k\left(\kappa_k \gamma_k D_k\right)^3}{\left(\kappa_l \gamma_l D_l\right)^2}+{\varsigma_l \kappa_l \gamma_l D_l}\right)}_{b_2/2} f_l^2+\underbrace{\rho_2 \kappa_l \gamma_l D_l}_{b_1} \frac{1}{f_l} \\
	\text { s.t. } ~~ & 0 \leq f_l \leq \frac{ \kappa_l \gamma_l D_l f_k^{\text{max}}}{\kappa_k \gamma_k D_k},~ \forall k \in \mathcal{K} .
\end{eqnarray}
\end{subequations}
Note that $g_1\left(f_l\right)$ can be characterized as $g_1\left(f_l\right)=b_2/2 f_l^2+b_1 / f_l$.
The minimum value of $g_1\left(f_l\right)$ is obtained at its stationary point. Thus, the optimal solution of problem (\ref{proof_4_3}) is
\begin{align}
		  	\label{proof_so_2}
	f_l^*=\min \left\{\sqrt[3]{\frac{b_1}{b_2}}, \min _{k \in \mathcal{K}} \frac{\kappa_l \gamma_l D_l f_l^{\max }}{\kappa_k \gamma_k D_k}\right\}
\end{align}
Combining (\ref{proof_so_1}) and (\ref{proof_so_2}), we obtain the desired result.

\section{Proof of Lemma \ref{Lemma 5}}
\label{proof_of_lemma_5}
Given $\boldsymbol{a}^*$ and $\chi^*$, problem (\ref{SP2}) can be transformed to
\begin{subequations}
	\label{proof_SP2-1}
	\begin{eqnarray}
		\label{SP2-1-function}
		& \underset{\boldsymbol{p}}{\min}   & \sum
		\limits_{n \in \mathcal{R}}a_{k,n}^*\frac{\rho_1(\xi_d+\xi_M)p_k}{W^{U} \log _2(1+\frac{h_k p_k^*}{I_n+W^{U} N_0})}  \\
		\label{chi_sp2}
		&\operatorname{s.t.} &\sum
		\limits_{n \in \mathcal{R}}a_{k,n}^*\frac{(\xi_d+\xi_M)}{W^{U} \log _2(1+\frac{h_k p_k^*}{I_n+W^{U} N_0})} \leq {\chi^*},  \\
		\label{p_sp2}
		&&  (\ref{p_k}).
	\end{eqnarray}
\end{subequations}
If $\sum_{n \in R} a_{k, n}^*=0$, then $p_k^*=0$. If the constraints in (\ref{proof_SP2-1}) are mutually contradictory, i.e.
\begin{align}
	p_k^{\max }<\frac{\left(I_{n_k^*}+W^{U} N_0\right)\left(2^{\frac{\xi_d+\xi_M}{W^{U} \chi^*}-1}\right)}{h_k},
\end{align}
which gives (\ref{TransDelay_hold}).
By eliminating $\rho_1$, $\xi_d$, $\xi_M$, and $W^{U}$, and denoting $\hat{p}_k=\frac{h_k p_k}{I_{n_k^*}+W^{U} N_0}$, (\ref{proof_SP2-1}) is  transformed to
\begin{subequations}
	\begin{eqnarray}
	& \underset{\hat{p}_k}{\min} & g_2\left(\hat{p}_k\right)=\frac{\hat{p}_k}{\log _2\left(1+\hat{p}_k\right)} \\
	&\operatorname{s.t.} & 0 \leq \hat{p}_k \leq \frac{h_k p_k^{\max }}{I_{n_k^*}+W^{U} N_0}, \\
	&& 2^{\frac{\xi_d+\xi_M}{W^{U} \chi^*}-1}-1 \leq \hat{p}_k.
	\end{eqnarray}
\end{subequations}
Thus we have
\begin{align}
	g_2^{\prime}\left(\hat{p}_k\right)=\frac{\log _2\left(1+\hat{p}_k\right)-\hat{p}_k/\left(\left(1+\hat{p}_k\right) \ln 2\right)}{\left(\log _2\left(1+\hat{p}_k\right)\right)^2}.
\end{align}
Then we can find that $g_2\left(\hat{p}_k\right)$ is monotonically increasing for $\hat{p}_k>0$. Thus, it can be obtained that
\begin{align}
	p_k^*=	\frac{\left(I_{n_k^*}+W^{U} N_0\right)\left(2^{\frac{\xi_d+\xi_M}{W^{U} \chi^*}-1}\right)}{h_k},
\end{align}
which completes the proof.

\section{Proof of Lemma \ref{Lemma 6}}
\label{proof_of_lemma_6}
Let $l$ denote the straggler among all SUs.
By fixing $\boldsymbol{a}^*$ and eliminating $\xi_d$ and $\xi_M$, problem (\ref{SP2}) is characterized as
\begin{subequations}
	\label{proof_SP2-3}
	\begin{eqnarray}
		\nonumber
		&\underset{\{{\boldsymbol{p}_k}\}_{k\in\mathcal{K}^*}}{\min} &\rho_1 \sum\limits_{n \in \mathcal{R}}\frac{p_k}{W^{U} \log _2(1+\frac{h_k p_k}{I_n+W^{U} N_0})}  + \rho_2 \sum\limits_{n \in \mathcal{R}}\frac{1}{W^{U} \log _2(1+\frac{h_k p_k}{I_n+W^{U} N_0})}  \\
		&\operatorname{s.t.}& \frac{\left(I_{n_k^*}+W^{U} N_0\right) h_l}{\left(I_{n_l^*}+W^{U} N_0\right) h_k} p_l \leq p_k,  \quad \forall k \in \mathcal{K}^* / l, \\
		&&0 \leq p_k \leq p_k^{\text{max}}, ~\forall k\in\mathcal{K}^*.
	\end{eqnarray}
\end{subequations}
Then we can obtain the optimal solution $p_k^*$ of (\ref{proof_SP2-3}) via solving the following decomposed problem
\begin{subequations}
	\label{proof_g2}
	\begin{eqnarray}
&\underset{\hat{p}_k}{\min} & g_2\left(\hat{p}_k\right)=\frac{\hat{p}_k}{\log _2\left(1+\hat{p}_k\right)} \\
&\operatorname{s.t.}& 0 \leq \hat{p}_k \leq \frac{h_k p_k^{\max }}{I_{n_k^*}+W^{U} N_0}, \\
&& \frac{h_l}{I_{n_l^*}+W^U N_0} p_l \leq \hat{p}_k.
\end{eqnarray}
\end{subequations}
Note that $g_2\left(\hat{p}_k\right)$ is monotonically increasing for $\hat{p}_k>0$. Thus, 
if $p_l\leq \frac{h_k p_k^{\text{max}}\left(I_{n_l^*}+W^U N_0\right)}{h_l\left(l_{n_k^*}+W^U N_0\right)}$,
we can obtain
\begin{align}
\label{lemma6-solution_1}
	p_k^*=\frac{h_l\left(I_{n_k^*}+W^U N_0\right)}{h_k\left(I_{n_l^*}+W^U N_0\right)} p_l, \quad \forall k \in \mathcal{K}^* / l.
\end{align}
Similar to (\ref{proof_g2}), let $\hat{p}_l=\frac{h_l p_l}{I_{n_l^*}+W^U N_0}$ and institute $p_k=p_k^*$ in (\ref{proof_SP2-3}) where $p_k=\frac{\left(I_{n_k^*}+W^U N_0\right) \hat{p}_l}{h_k}$ ), 
we obtain the following problem regarding $\hat{p}_l$
\begin{subequations}
	\label{proof_lemma6_final}
	\begin{eqnarray}
&\underset{\hat{p}_l}{\min} & g_3\left(\hat{p}_l\right)= \rho_1 \underbrace{ \sum_{k\in\mathcal{ K }^*} \frac{I_{n_k^*}+W^U N_0}{h_k}}_{c_1} \frac{\hat{p}_l}{\log _2\left(1+\hat{p}_l\right)}+\frac{\rho_2}{\log _2\left(1+\hat{p}_l\right)}  \\
&\operatorname{s.t.}& 0\leq \hat{p}_l\leq \frac{h_k p_k^{\text{max}}}{I_{n_k^*}+W^UN_0}, ~\forall k \in\mathcal{K}^*.
\end{eqnarray}
\end{subequations}
Note that $g_3\left(\hat{p}_l\right)$ can be rewritten as $g_3\left(\hat{p}_l\right)=\frac{\rho_1c_1 \hat{p}_l}{\log _2\left(1+\hat{p}_l\right)}+\frac{\rho_2}{\log _2\left(1+\hat{p}_l\right)}$.
Thus we can find that $g_3\left(\hat{p}_l\right)$ has a unique minimum point 
$\hat{p}_l^0 \in\left(0, c_2\right]$ where $c_2=2^{\left(1+\sqrt{\max \left\{\frac{\rho_2}{\rho_1c_1}, 1\right\}-1}\right) / \ln 2}$ such that $g_3^{\prime}\left(\hat{p}_l^0\right)=0$.
Hence, the optimal solution of the problem (\ref{proof_lemma6_final}) can be expressed as
\begin{align}
\label{lemma6-solution_2}
\hat{p}_l^*=	{\min}\left\lbrace \underset{k\in\mathcal{K}^*}{\min}\frac{h_k p_k^{max}}{I_{n_k^*}+W^{U} N_0},\hat{p}_l^0\right\rbrace.
\end{align}
By combining (\ref{lemma6-solution_1}) and (\ref{lemma6-solution_2}), we complete the proof.

\end{document}